\documentclass[aps,prd,showpacs,twocolumn,superscriptaddress,nofootinbib,preprintnumbers]{revtex4-1} 

\makeatletter
\def\p@subsection{}
\makeatother
\usepackage{color,graphicx,amsmath,amssymb,lipsum,bm}
\usepackage[colorlinks=true,citecolor=blue,linkcolor=blue,urlcolor=blue, backref=false,pdfborder={0 0 0}]{hyperref}
\usepackage[normalem]{ulem}
\usepackage[utf8]{inputenc} 
\usepackage{mathtools}
\usepackage{float}
\usepackage{ulem}
\usepackage{diagbox}
\usepackage[dvipsnames]{xcolor}
\usepackage{mathrsfs}
\usepackage{multirow}
\usepackage{amsmath}

\newcommand{\be}{\begin{equation}}
\newcommand{\ee}{\end{equation}}
\newcommand{\beqa}{\begin{eqnarray}}
\newcommand{\eeqa}{\end{eqnarray}}

\newcommand{\bseq}{\begin{subequations}}
\newcommand{\eseq}{\end{subequations}}

\newcommand{\di}{\mathrm d}

\def\gsim{\raise0.3ex\hbox{$\;>$\kern-0.75em\raise-1.1ex\hbox{$\sim\;$}}}
\def\lsim{\raise0.3ex\hbox{$\;<$\kern-0.75em\raise-1.1ex\hbox{$\sim\;$}}}
\def\beqn#1{\begin{equation}\label{#1}}
\def\eeqn{\end{equation}}
\def\beqa#1{\begin{eqnarray}\label{#1}}
\def\eeqa{\end{eqnarray}}

\def\Z2{$\mathcal{Z_2}$}

\newcommand {\ignore}[1]{}

\renewcommand{\arraystretch}{1.3}

\begin{document}

\title{Constraining Dynamical Dark Energy with Tomographic Galaxy and \\CMB Weak Lensing Through Higher-Order Statistics}

\author{Shu-Fan Chen} 
\email{sc5848@columbia.edu
 } 
\affiliation{Astrophysics Laboratory, Columbia University, New York, NY 10027, USA}
\affiliation{Department of Physics, Columbia University New York, NY 10027, USA}
\affiliation{Department of Astronomy, Columbia University New York, NY 10027, USA}

\author{J.~Colin Hill}
\affiliation{Astrophysics Laboratory, Columbia University, New York, NY 10027, USA}
\affiliation{Department of Physics, Columbia University New York, NY 10027, USA}

\author{Zolt\'an Haiman}
\affiliation{Institute of Science and Technology Austria (ISTA), Am Campus 1, Klosterneuburg 3400, Austria}
\affiliation{Department of Astronomy, Columbia University New York, NY 10027, USA}
\affiliation{Department of Physics, Columbia University New York, NY 10027, USA}

\begin{abstract}
Tomographic weak gravitational lensing probes cosmic structure growth and geometry, making it a powerful probe of dynamical dark energy. We forecast information beyond the lensing power spectrum from non-Gaussian statistics of tomographic galaxy and CMB lensing convergence maps. Using the $\texttt{DDELenS}$ suite of ray-traced $N$-body simulations, we build Gaussian process emulators for power spectra, Minkowski functionals, moments, peak and minima counts, and the wavelet scattering transform, varying $\Omega_m$, $\sigma_8$, $w_0$, and $w_a$. We consider Rubin-like galaxy lensing and Simons Observatory-like CMB lensing over $15{,}000\,\mathrm{deg}^2$, and validate the analysis using a withheld fiducial cosmology, analytical power-spectrum forecasts, and Gaussian random-field null tests. As an initial step, we neglect baryonic effects and intrinsic alignments and approximate galaxy sources as discrete redshift planes. Using all source redshifts at $z_s=\{0.5, 1, 1.5, 2, 2.5, 1100\}$, combining all five non-Gaussian statistics reduces uncertainties on $w_0$ and $w_a$ by $24.3\%$ and $28.0\%$, and on $\Omega_m$ and $\sigma_8$ by $41.9\%$ and $45.8\%$, relative to the power-spectrum-only baseline. The resulting $w_0$ and $w_a$ uncertainties are $34.7\%$ and $48.7\%$ smaller than current DESI+DES constraints. If the DESI+DES best-fit values remain unchanged, this precision would increase the significance of the departure from $\Lambda$CDM from $3.0\sigma$ to $5.2\sigma$, compared with $3.8\sigma$ from the power spectrum alone, assuming a Gaussian posterior and unchanged $w_0$--$w_a$ correlation. Minkowski functionals and moments provide comparable constraining power, while combining them with the other summaries gives the strongest constraints. Broad galaxy-lensing redshift coverage is particularly important for dynamical dark energy, while CMB lensing provides complementary information.
\end{abstract}

\maketitle

\section{Introduction}
Understanding the physical origin of the accelerated expansion of the Universe remains a central goal of cosmology. In the standard $\Lambda$CDM model, dark energy is described by a cosmological constant with a time-independent equation of state, $w=-1$. Recent measurements from the Dark Energy Spectroscopic Instrument (DESI), including baryon acoustic oscillations (BAO)~\cite{DESI:2024mwx,DESI:2025zpo,DESI:2025zgx} and full-shape galaxy power spectrum measurements~\cite{DESI:2024jxi,DESI:2024hhd}, have increased interest in models with an evolving dark-energy equation of state. When combined with CMB measurements and Type Ia supernova data from Pantheon+, Union3, or DES~\cite{Brout:2022vxf,Rubin:2023jdq,DES:2024jxu,DES:2025sig}, these observations show a moderate preference for $w_0>-1$ and $w_a<0$ in the Chevallier--Polarski--Linder (CPL) parametrization~\cite{Chevallier:2000qy,Linder:2002et}. More recently, a joint analysis of supernovae, BAO, weak lensing, and galaxy clustering from the full Dark Energy Survey (DES) dataset reported a preference for dynamical dark energy ranging from $2.2\sigma$ to $3.0\sigma$, depending on the combination of external datasets~\cite{DES:2026jmi}. Whether these indications reflect genuine new physics or residual systematics in supernova and large-scale-structure measurements remains an open question~\cite{Gialamas:2024lyw,Dhawan:2024gqy,Chen:2024vuf,Chen:2025jnr}. Resolving this question will require continued scrutiny of existing datasets, quantitative forecasts of the additional discriminating power that future observations can provide, and the collection of new data with Rubin, Euclid, Roman, DESI, and other surveys.

Dynamical dark energy affects both the expansion history of the Universe and the growth of cosmic structure. Weak gravitational lensing is particularly useful in this context because it is sensitive both to the matter distribution along the line of sight and to the distances between the observer, lenses, and sources~\cite{LSSTScience:2009jmu,Lewis:2006fu}. Measurements at multiple source redshifts can therefore provide direct information about the redshift evolution of the lensing signal and help distinguish the effects of $w_0$ and $w_a$ from those of other cosmological parameters.

Galaxy weak-lensing measurements from DES~\cite{DES:2026mkc,DES:2026fyc}, the Kilo-Degree Survey (KiDS)~\cite{Fluri:2022rvb,Wright:2025xka}, and the Hyper Suprime-Cam Subaru Strategic Program (HSC)~\cite{Miyatake:2023njf} have already demonstrated the constraining power of cosmic shear and related lensing observables. The next generation of galaxy weak-lensing measurements from the Vera C. Rubin Observatory's Legacy Survey of Space and Time~\cite{LSSTDarkEnergyScience:2018yem}, the Nancy Grace Roman Space Telescope~\cite{Akeson:2019biv,Spergel:2015sza}, and Euclid~\cite{Euclid:2020rfv} will substantially increase both the survey area and the number of well-measured source galaxies. Dividing these sources into redshift bins provides tomographic information that helps separate the effects of cosmological parameters on the lensing signal~\cite{Liu:2018dsw,Li:2018owg,Coulton:2018ebd}. CMB lensing provides a complementary view of the matter distribution, with a source plane at the epoch of recombination and a broad lensing kernel extending beyond the redshift range most directly probed by galaxy lensing~\cite{Lewis:2006fu}. CMB lensing has already been measured at high signal-to-noise with \textit{Planck}~\cite{Planck:2018lbu,Carron:2022eyg}, ACT~\cite{ACT:2023dou,ACT:2023kun}, and SPT/SPT-3G~\cite{Wu:2019hek,SPT:2023jql}, while the Simons Observatory will provide substantially improved sensitivity~\cite{SimonsObservatory:2018koc,SimonsObservatory:2025wwn}. The overlap between galaxy and CMB lensing in the large-scale structure they trace produces correlations that can provide additional cosmological information and reduce parameter degeneracies~\cite{Chen:2026zqq}.

The power spectrum is a natural starting point for extracting cosmological information from these observations. For a Gaussian field, it provides a complete statistical description, but nonlinear gravitational evolution generates non-Gaussian structure that is not fully captured by two-point statistics. Lensing maps can therefore contain additional information in their amplitude distributions, spatial morphology, and higher-order correlations. Several summary statistics have been developed to access this information, including Minkowski functionals and moments~\cite{Petri:2013ffb,Petri:2015ura,Marques:2018ctl}, peak and minima counts~\cite{Dietrich:2009jq,Li:2018owg,Coulton:2019enn}, and the wavelet scattering transform~\cite{Cheng:2020qbx,Cheng:2021xdw,Boone:2025obt}. For dynamical dark energy, however, non-Gaussian information alone cannot replace the redshift leverage provided by lensing tomography. A single source plane provides only limited sensitivity to the evolution of the dark-energy equation of state, leaving strong degeneracies among $w_0$, $w_a$, and other cosmological parameters. Combining multiple source redshifts and their cross-correlations is therefore particularly important for constraining dynamical dark energy. The relevant question is then how much additional information non-Gaussian statistics provide once this tomographic information is included. CMB lensing can further contribute by extending the redshift baseline and by constraining parameter combinations that are degenerate in galaxy lensing alone. We therefore separately quantify the gains from increasing the source-redshift coverage and from adding non-Gaussian statistics, and examine how these two sources of information complement one another.

In this work, we use the {\tt DDELenS} simulation suite~\cite{Chen:2026zqq} to forecast constraints on $\Omega_m$, $\sigma_8$, $w_0$, and $w_a$ from tomographic galaxy and CMB lensing. We adopt a Rubin-like galaxy lensing survey and Simons Observatory-like CMB lensing as a concrete case study --- though the methodology applies equally to Roman and Euclid --- and compare several combinations of source redshifts. Using Gaussian process emulators trained on ray-traced $N$-body simulations, we predict the power spectra and five sets of additional summary statistics: Minkowski functionals, moments, peak counts, minima counts, and the wavelet scattering transform. We quantify the improvement relative to the power-spectrum baseline when these statistics are included individually or jointly. We also examine the contributions of individual Minkowski functionals and moments to identify which components drive their constraining power. For these initial forecasts, we neglect baryonic effects and intrinsic alignments and approximate galaxy source populations as discrete redshift planes, allowing us to assess the statistical potential of the summary statistics within a simplified experimental setup.  We will revisit these effects in future work.

The remainder of this paper is organized as follows. Section~\ref{sec:simulation} describes the simulations and convergence maps. Section~\ref{sec:method} presents the experimental configurations, summary statistics, emulators, and likelihood. Section~\ref{sec:validation_and_null_tests} describes the validation and null tests. Section~\ref{sec:results} presents the forecast constraints and examines the contributions of the different statistics and source-redshift combinations. We conclude in Section~\ref{sec:conclusions}.

\section{Simulations}\label{sec:simulation}
We use the {\tt DDELenS} $N$-body simulation suite~\cite{Chen:2026zqq} to generate fully ray-traced\footnote{Note that using the Born approximation in the construction of convergence maps can lead to biased parameter estimates with higher-order statistics for next-generation weak lensing surveys~\cite{Petri:2016qya,Fabbian:2017wfp,Pratten:2016dsm}.} lensing convergence maps for the configurations considered in this work. The suite follows $1024^3$ particles in a periodic box of side length $512\,h^{-1}\mathrm{Mpc}$ and incorporates dynamical dark energy using a linear-species approximation through an $N$-body-gauge implementation using {\tt PKDGRAV3} and {\tt CONCEPT}~\cite{Potter:2016ttn,Dakin:2019dxu,Dakin:2019vnj}. We summarize the map construction relevant to this analysis below; the simulation pipeline and its validation are described in Ref.~\cite{Chen:2026zqq}. The suite consists of 128 simulations run with {\tt PKDGRAV3}~\cite{Potter:2016ttn}, sampling four cosmological parameters: $\Omega_m$, $\sigma_8$, $w_0$, and $w_a$ using a Sobol sequence over the following simulation-domain ranges:
\begin{alignat}{2}
    &\Omega_m \in [0.28,\,0.35]\,,\quad &\sigma_8 \in [0.74,\,0.90]\,, \nonumber\\
    &w_0 \in [-1.0,\,-0.5]\,,\quad &w_a \in [-1.5,\,0.0]\,.
\end{alignat}
This simulation domain is motivated by the DESI DR1 and DR2 results considered in the suite paper~\cite{DESI:2024hhd,DESI:2025zgx}. The fiducial $\Lambda$CDM cosmology used to estimate the covariance matrix has $\Omega_m = 0.300$, $\sigma_8 = 0.8117$, $w_0 = -1$, and $w_a = 0$. We set the fiducial cosmology for sampling to be at $\Omega_m =0.315$, $\sigma_8=0.82$, $w_0=-0.75$, $w_a=-0.75$ instead so that it is located at the center of the simulated cosmologies. All 128 cosmologies share the same initial random seed, reducing sample-variance differences between cosmologies. For each cosmology and source redshift, we generate 10,000 lensing convergence maps with {\tt LensTools}~\cite{Petri:2016zrq}, using random slicing, shifts, and orientations. At the fiducial $\Lambda$CDM cosmology, we run simulations with two additional sets of initial conditions. We combine the three sets to generate 100,000 lensing convergence maps for covariance estimation.

Each map spans $3.5\,\mathrm{deg}$ per side and contains $512\times512$ pixels. To forecast constraints for future surveys, we rescale the covariance estimated from these maps to a survey area of approximately $15,000\,\mathrm{deg}^2$. We restrict the map modes to $\ell\in[300,3000]$, using a smooth taper based on the normalized bump cumulative distribution at the boundaries to suppress ringing in real space (see Ref.~\cite{Chen:2026jlr}). We also apply a Wiener filter, $C_\ell^{\rm signal}/(C_\ell^{\rm signal}+N_{\ell})$, in Fourier space to enhance the signal relative to the noise (see Figure~\ref{fig:cl_nl}). Here $C_\ell^{\rm signal}$ is the theoretical lensing power spectrum computed from {\tt PYCCL}~\cite{LSSTDarkEnergyScience:2018yem}\footnote{Code and documentation can be found in \url{https://github.com/LSSTDESC/CCL} and \url{https://ccl.readthedocs.io/en/latest/index.html}, respectively.} and $N_\ell$ is the corresponding noise power spectrum. The map-generation strategy builds on that used for {\tt MassiveNuS}~\cite{Liu:2017now,Petri:2016zrq}. In {\tt DDELenS}, matched lens-plane selections and shifts preserve the shared line-of-sight structure across source redshifts, enabling the tomographic auto- and cross-spectra used here.

\section{Methods}\label{sec:method}
\subsection{Experimental Setup}
We consider two experimental configurations. For low-redshift galaxy weak lensing, we adopt a Rubin-like survey with the projected galaxy number densities listed in Table~\ref{tab:rubin_projected_density} \cite{LSSTScience:2009jmu,LSSTDarkEnergyScience:2018yem,Coulton:2018ebd}. We add Gaussian noise with variance $\sigma_{\rm shape}^2/\bar{n}_g$, where $\sigma_{\rm shape}=0.3$ is the intrinsic shape noise. In this initial forecast, we neglect baryonic effects and intrinsic alignments, leaving their impact to future work. We also approximate each source population as a plane at a single redshift.

\begin{figure}
    \centering
    \includegraphics[width=0.95\linewidth]{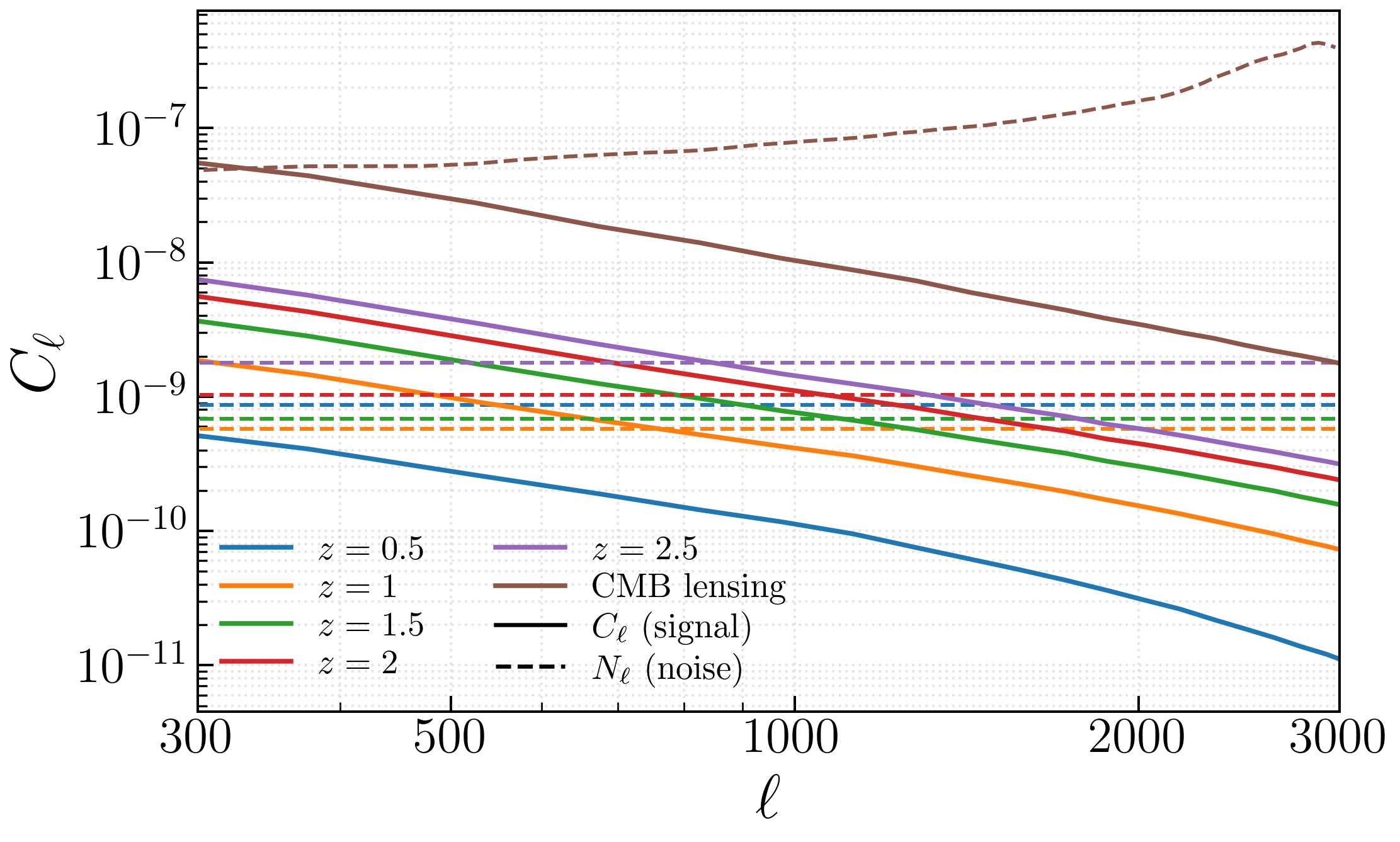}
    \caption{Lensing convergence power spectra of the signal (solid) and noise (dashed) for each galaxy source redshift and for CMB lensing. The signal is the mean over noiseless realizations at the fiducial cosmology from {\tt DDELenS}. The galaxy noise is the shape noise, set by each bin's galaxy number density as described in Table~\ref{tab:rubin_projected_density} from Rubin-like galaxy survey; the CMB lensing noise is the nominal SO quadratic-estimator lensing reconstruction noise.  The multipole range plotted here is that used in our forecasts; additional information will be present on larger scales (lower $\ell$) in the actual survey data.}
    \label{fig:cl_nl}
\end{figure}

For CMB lensing, we follow the same approach as in Ref.~\cite{Chen:2026jlr}. We adopt nominal Simons Observatory (SO) noise from Ref.~\cite{SimonsObservatory:2018koc}\footnote{\url{https://github.com/simonsobs/so_noise_models}} after standard internal linear combination (ILC) component separation of the primary CMB\footnote{This is a more conservative setup than the enhanced SO Large Aperture Telescope survey described in Ref.~\cite{SimonsObservatory:2025wwn}; thus our CMB lensing forecasts are conservative in this regard.}.  We then forward-model the lensing reconstruction noise using the quadratic estimator \cite{Hu:2001kj,Madhavacheril:2020ido}. We combine all available estimators, $\{TT,TE,EE,TB,EB\}$, to construct the minimum-variance quadratic estimator. The simulation and reconstruction pipeline is implemented with {\tt pixell}\footnote{\url{https://github.com/simonsobs/pixell}} and {\tt symlens}\footnote{\url{https://github.com/simonsobs/symlens}}. We hold the cosmology used to generate the unlensed CMB fixed in order to isolate the information carried by the lensing convergence field. Figure~\ref{fig:cl_nl} compares the resulting noise power spectra with the signal for each galaxy source plane and for CMB lensing.

We assume the same $15,000$ deg$^2$ sky coverage for both galaxy and CMB lensing, with complete overlap between the survey footprints. This common coverage allows us to analyze the convergence fields jointly and include their cross-spectra alongside the auto-spectra. We consider several combinations of galaxy source planes, both with and without CMB lensing, to assess how the source-redshift coverage affects the parameter constraints and the additional information provided by non-Gaussian statistics.

\begin{table}[]
    \centering
    \begin{tabular}{c|ccccc}
        Source Redshift $z_s$ & 0.5 & 1.0 & 1.5 & 2.0 & 2.5 \\
        \hline
        $\bar{n}_g$ $[{\rm arcmin}^{-2}]$ & 8.83 & 13.25 & 11.15 & 7.36& 4.26
    \end{tabular}
    \caption{Projected galaxy number densities $\bar{n}_g$ for the source planes used in the Rubin-like survey forecast. The source redshifts are denoted by $z_s$.}
    \label{tab:rubin_projected_density}
\end{table}

\subsection{Summary Statistics}
Following Ref.~\cite{Chen:2026jlr}, we consider several statistics that probe information beyond the power spectrum: Minkowski functionals (MFs), moments (MOM), peak counts (PC), minima counts (MC), and the wavelet scattering transform (WST). We use the power spectra as the baseline for our forecasts and assess the additional constraining power of each summary-statistic set by combining it with this baseline. We also analyze the joint combination of all the statistics to determine how much their combined use improves the constraints. Below, we describe the definitions and measurement choices for each statistic. Note that before computing all statistics, maps are already filtered to consider only modes $\ell\in[300,3000]$ as discussed in Section~\ref{sec:simulation}.  For real-space statistics, we also apply a Wiener filter based on Figure~\ref{fig:cl_nl} in Fourier space prior to measurement.

\subsubsection{Power spectrum}
We measure the power spectrum in bins of width $\Delta\ell=150$ using the estimator
\begin{align}
    C^{XY}_{\ell_i} = \frac{1}{N_i}\sum_{\boldsymbol{\ell}\in\ell_i}\kappa^X(\boldsymbol{\ell})\kappa^Y(-\boldsymbol{\ell})\,,
\end{align}
where $X$ and $Y$ label the source redshift bins, $\ell_i$ denotes the $i$th multipole bin, and $N_i$ is the number of modes in that bin. Because the convergence maps are correlated across source redshifts, we include all corresponding auto- and cross-spectra.

\subsubsection{Minkowski functionals}
We consider all the three Minkowski functionals (MFs) available for a two-dimensional field~\cite{Minkowski1903,Chen:2026jlr}: $V_0$, $V_1$, and $V_2$. They are defined as
\begin{align}
    V_0(\nu) &= \frac{1}{A}\int_{\Sigma(\nu)} \di a\,,\nonumber\\
    V_1(\nu) &= \frac{1}{4A}\int_{\partial\Sigma(\nu)} \di\ell\,,\nonumber\\
    V_2(\nu) &= \frac{1}{4\pi A}\int_{\partial\Sigma(\nu)}\mathcal{K}\di\ell\,,
\end{align}
where $A$ is the map area, $\Sigma(\nu)$ is the excursion set above threshold $\nu$, $\partial\Sigma(\nu)$ is its boundary, and $\mathcal{K}$ is the local boundary curvature. For each MF, we use nine equally spaced threshold bins between $-3\sigma_{\rm fid}$ and $3\sigma_{\rm fid}$, where $\sigma_{\rm fid}$ is the convergence map rms computed at the fiducial cosmology\footnote{This is a conservative choice motivated by our previous work on CMB lensing~\cite{Chen:2026jlr}. Additional information may be present beyond the $\pm3\sigma$ range, particularly for low-redshift galaxy lensing, where the convergence field is more non-Gaussian.  We leave exploration of this choice for future work.}.

\subsubsection{Moments}
We consider four moments: $\sigma_0$, $\sigma_1$, $S_1$, and $K_1$, defined as
\begin{alignat}{2}
    &\sigma_0 \equiv \sqrt{\langle\kappa^2\rangle}\,,\quad &\sigma_1 &\equiv \sqrt{\langle|\nabla\kappa|^2\rangle}\,,\nonumber\\
    &S_1 \equiv \langle\kappa^2\nabla^2\kappa\rangle\,,\quad &K_1 &\equiv \langle\kappa^3\nabla^2\kappa\rangle\,.
\end{alignat}
The first two quantities, $\sigma_0$ and $\sigma_1$, characterize the root-mean-square convergence amplitude and gradient amplitude, respectively. The higher-order moments, $S_1$ and $K_1$, combine powers of the convergence with its Laplacian. This choice follows our earlier CMB-lensing analysis and is motivated by perturbative descriptions of Minkowski functionals~\cite{Chen:2026jlr,Petri:2013ffb,Matsubara:2000mw,Matsubara:2010te}. Thus, all the moments except $\sigma_0$ involve spatial derivatives of the field. In addition to analyzing these four quantities jointly, we examine their individual contributions to the parameter constraints to identify which moments account for the improvement over the power-spectrum baseline.

\subsubsection{Peak/Minima Counts}
Peak and minima counts measure the abundance of local maxima and minima as a function of their amplitudes~\cite{Bardeen:1985tr,1987MNRAS.226..655B,Chen:2026jlr}. We analyze these two statistics separately to assess the constraining power of the high- and low-amplitude extrema in the convergence field. We use six equally spaced amplitude bins for each statistic, spanning $[-\sigma_{\rm fid},3\sigma_{\rm fid}]$ for peak counts and $[-3\sigma_{\rm fid},\sigma_{\rm fid}]$ for minima counts. As for the Minkowski functionals, the bin ranges are defined using the fiducial value $\sigma_{\rm fid}$, providing a common amplitude scale across the sampled cosmologies.

\subsubsection{Wavelet Scattering Transform}
The WST uses successive wavelet convolutions and modulus operations to summarize multiscale structure beyond two-point correlations~\cite{2011arXiv1101.2286M,Cheng:2021xdw}. Its use for weak lensing has been explored in Refs.~\cite{Cheng:2020qbx,Boone:2025obt}. Following our CMB-lensing analysis~\cite{Chen:2026jlr}, we use Morlet wavelets (see Ref.~\cite{2012arXiv1203.1513B} for more details) to compute the WST. For our $512\times512$ maps, we set $J=8$ and $L=1$. We perform the measurements with {\tt PyWST}~\cite{Regaldo-SaintBlancard:2020dlb}\footnote{The code is available at \url{https://github.com/bregaldo/pywst}.}.

\subsection{Gaussian Process Emulation}
Following the emulator framework of Ref.~\cite{Chen:2026jlr}, we train Gaussian process (GP) emulators~\cite{2006gpml.book.....R} for the power spectra and additional summary statistics using simulations at the sampled cosmologies. We adopt a Mat{\'e}rn kernel~\cite{matern2013spatial} with an additional white-noise kernel whose variance is estimated from 10,000 realizations at the fiducial cosmology. We take the $\nu\rightarrow\infty$ limit of the Mat{\'e}rn kernel, yielding an infinitely differentiable kernel.  Validation of the emulators is described in Section~\ref{sec:validation_and_null_tests}.

\begin{figure*}
    \centering
    \includegraphics[width=0.65\linewidth]{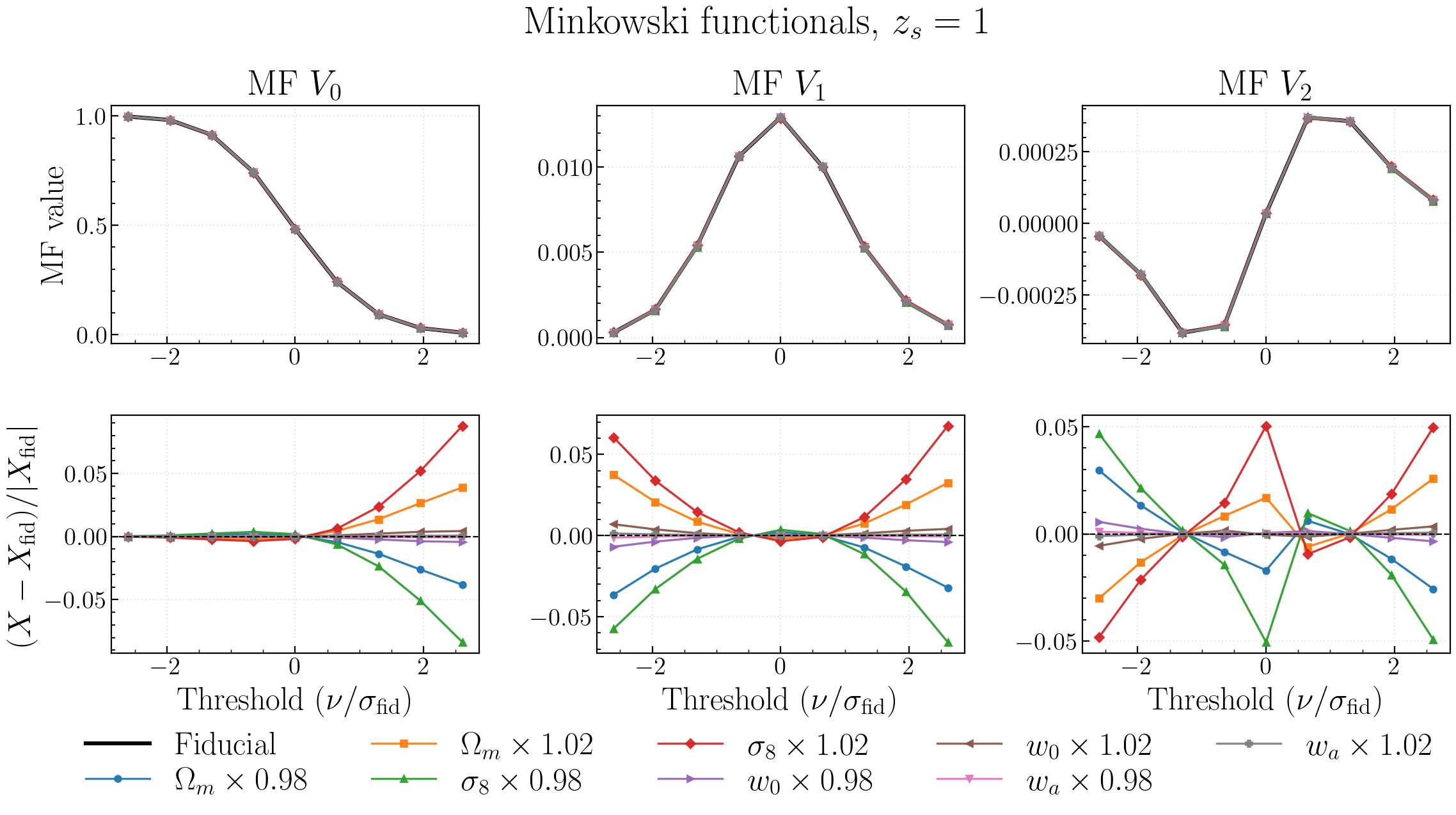}
    \par\vspace{0.5cm}
    \includegraphics[width=0.65\linewidth]{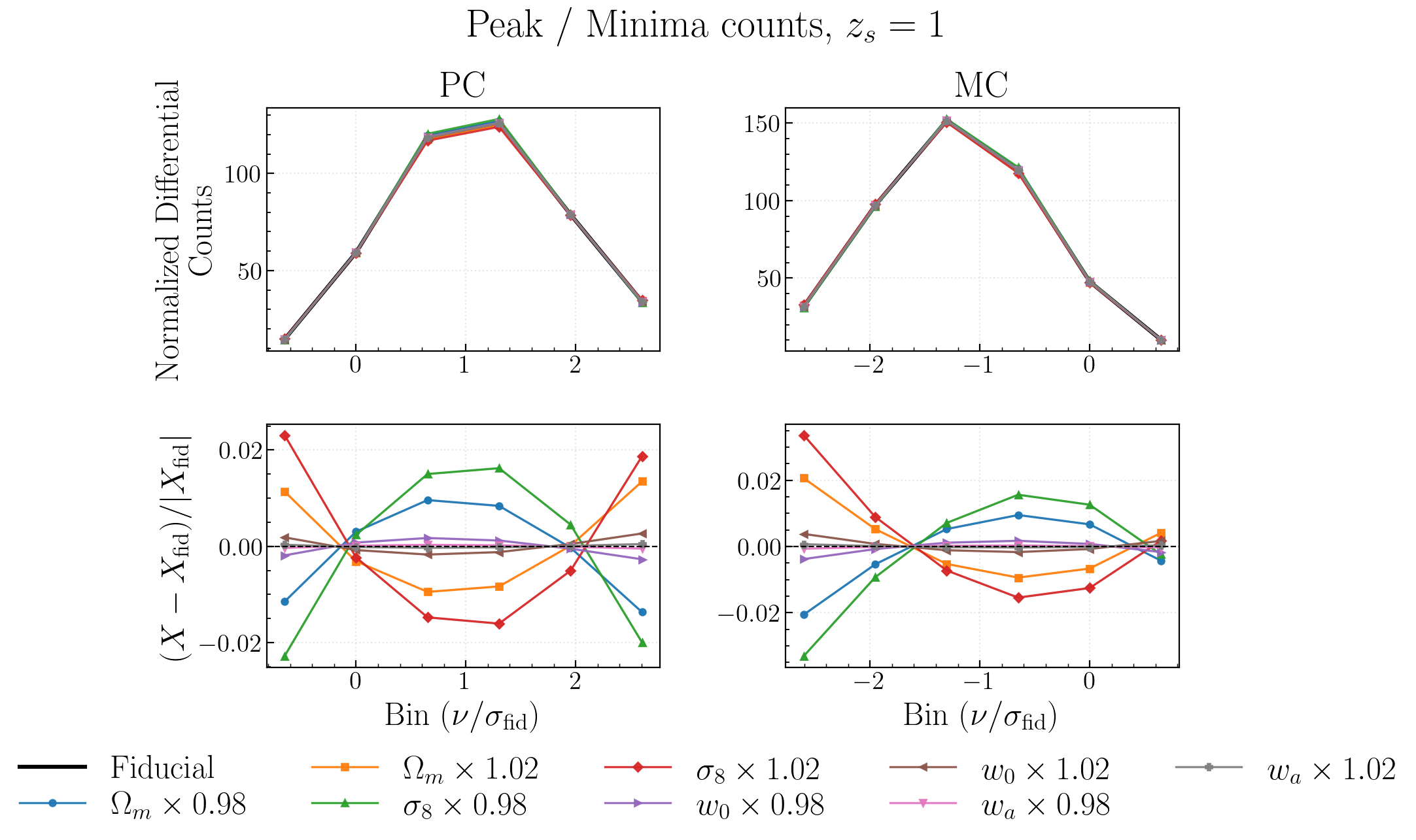}
    \par\vspace{0.5cm}
    \includegraphics[width=0.65\linewidth]{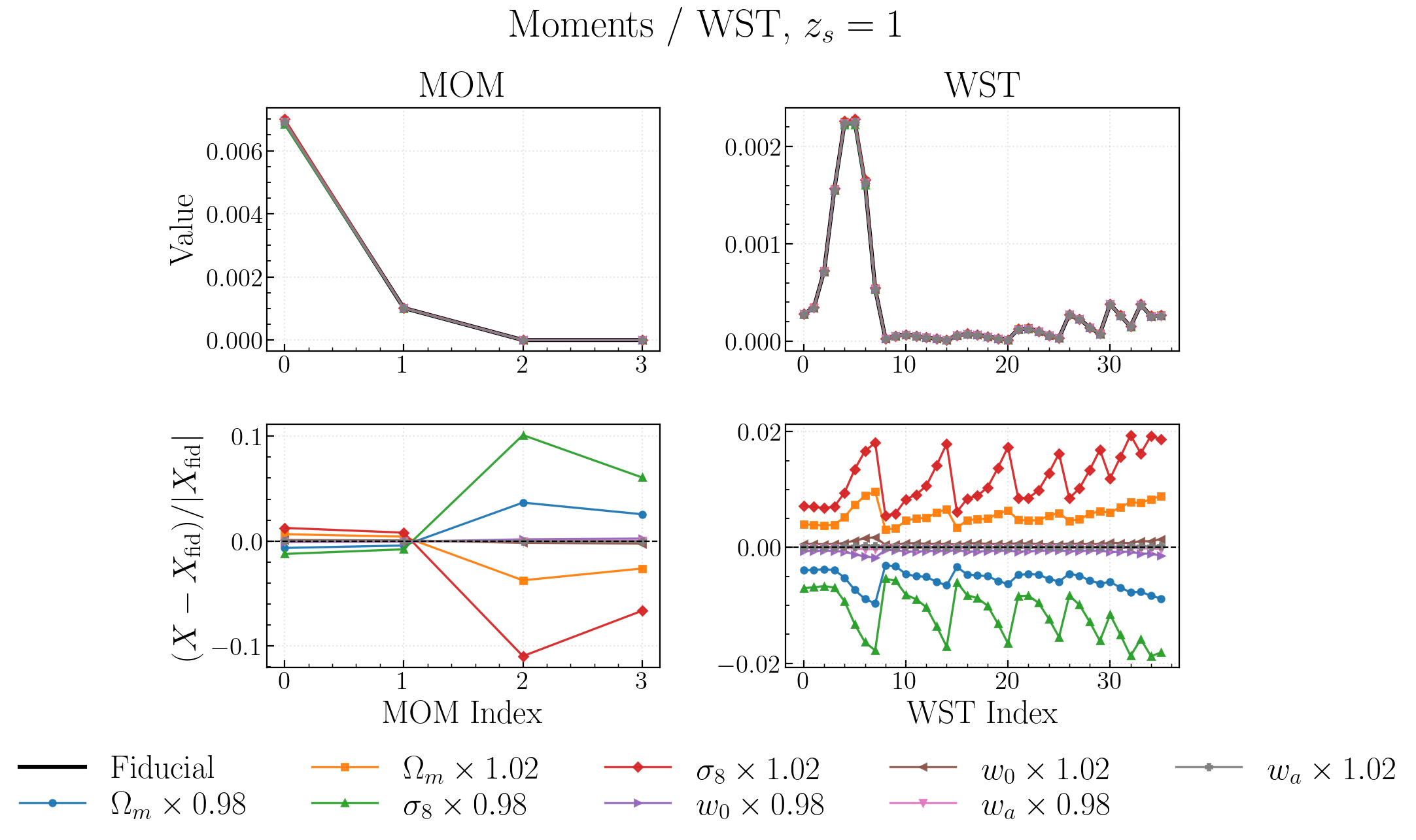}
    \caption{Non-Gaussian statistics and their fractional responses to changes in cosmological parameters at $z_s=1$. From top to bottom, the panels show the Minkowski functionals $V_0$, $V_1$, and $V_2$ as functions of the normalized threshold $\nu/\sigma_0$; peak counts (PC) and minima counts (MC); and the four moments ($\sigma_0$, $\sigma_1$, $S_1$, and $K_1$) and 36 WST coefficients ($J=8$, $L=1$). For each statistic, the upper row shows its value, with the fiducial cosmology in black, and the lower row shows the fractional response $(X-X_{\rm fid})/|X_{\rm fid}|$. Colored curves correspond to parameters rescaled by factors of $0.98$ and $1.02$, with all other parameters held fixed. All curves are emulator predictions.}
    \label{fig:fractional_z10}
\end{figure*}

\begin{figure*}
    \centering
    \includegraphics[width=0.65\linewidth]{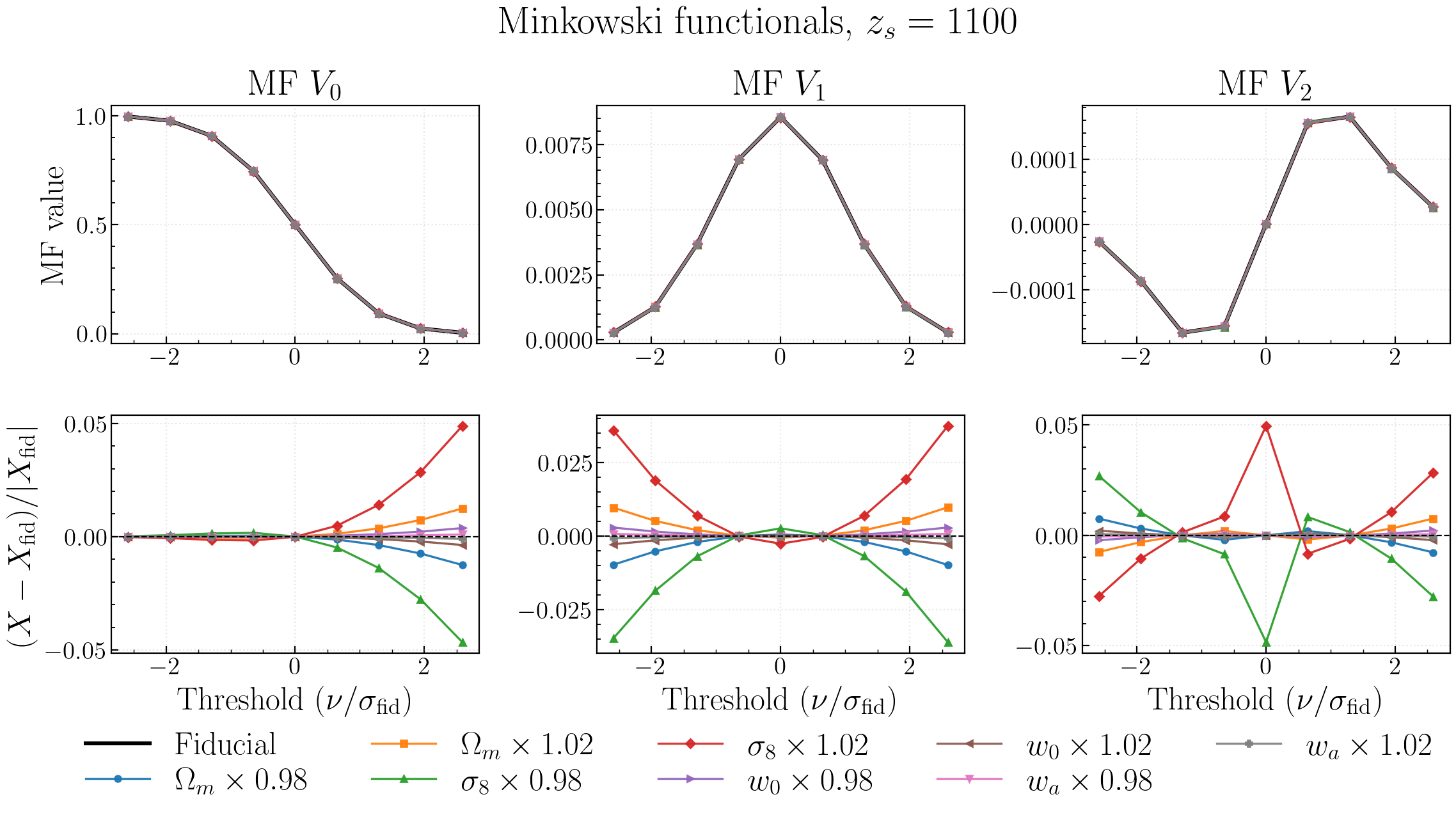}
    \par\vspace{0.5cm}
    \includegraphics[width=0.65\linewidth]{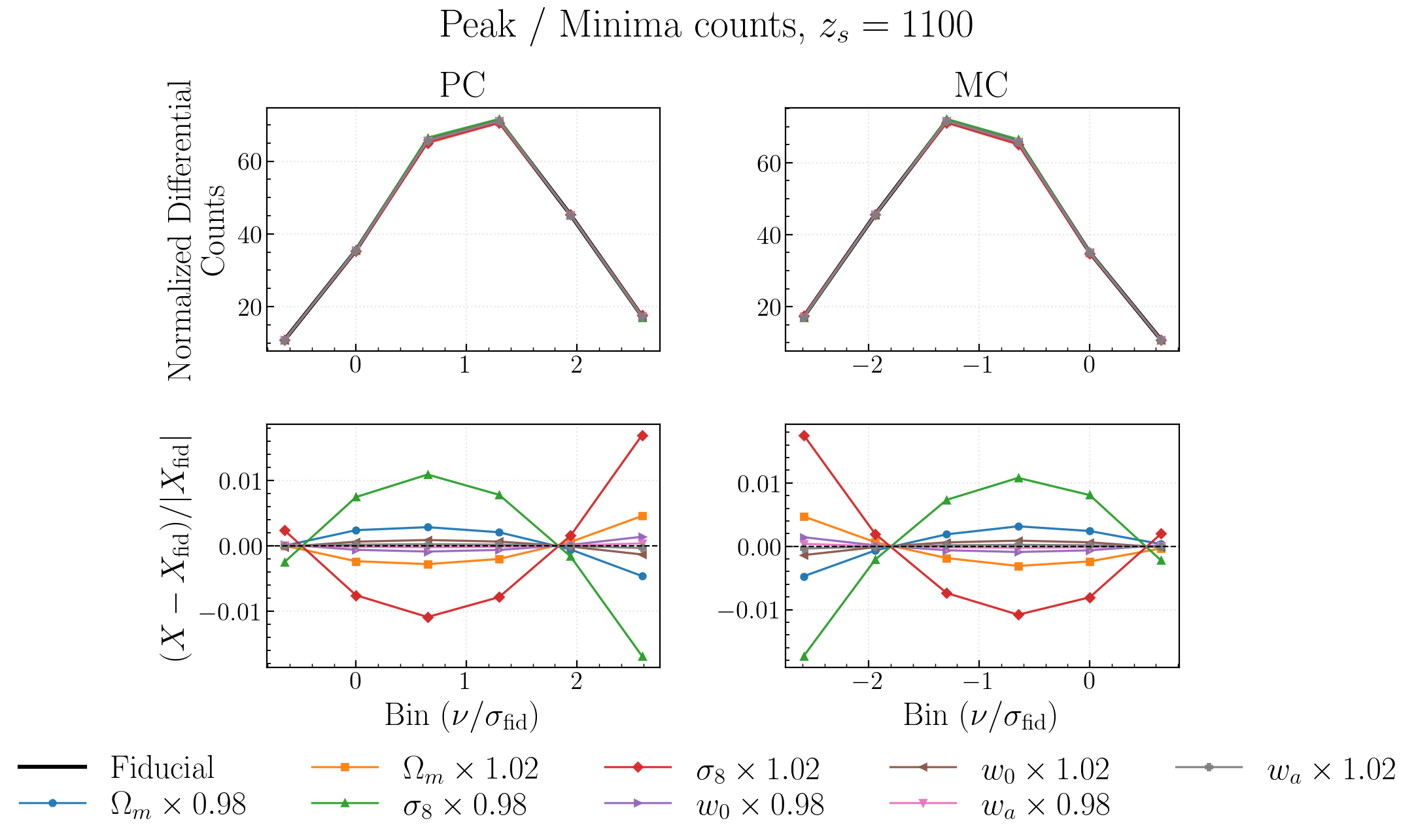}
    \par\vspace{0.5cm}
    \includegraphics[width=0.65\linewidth]{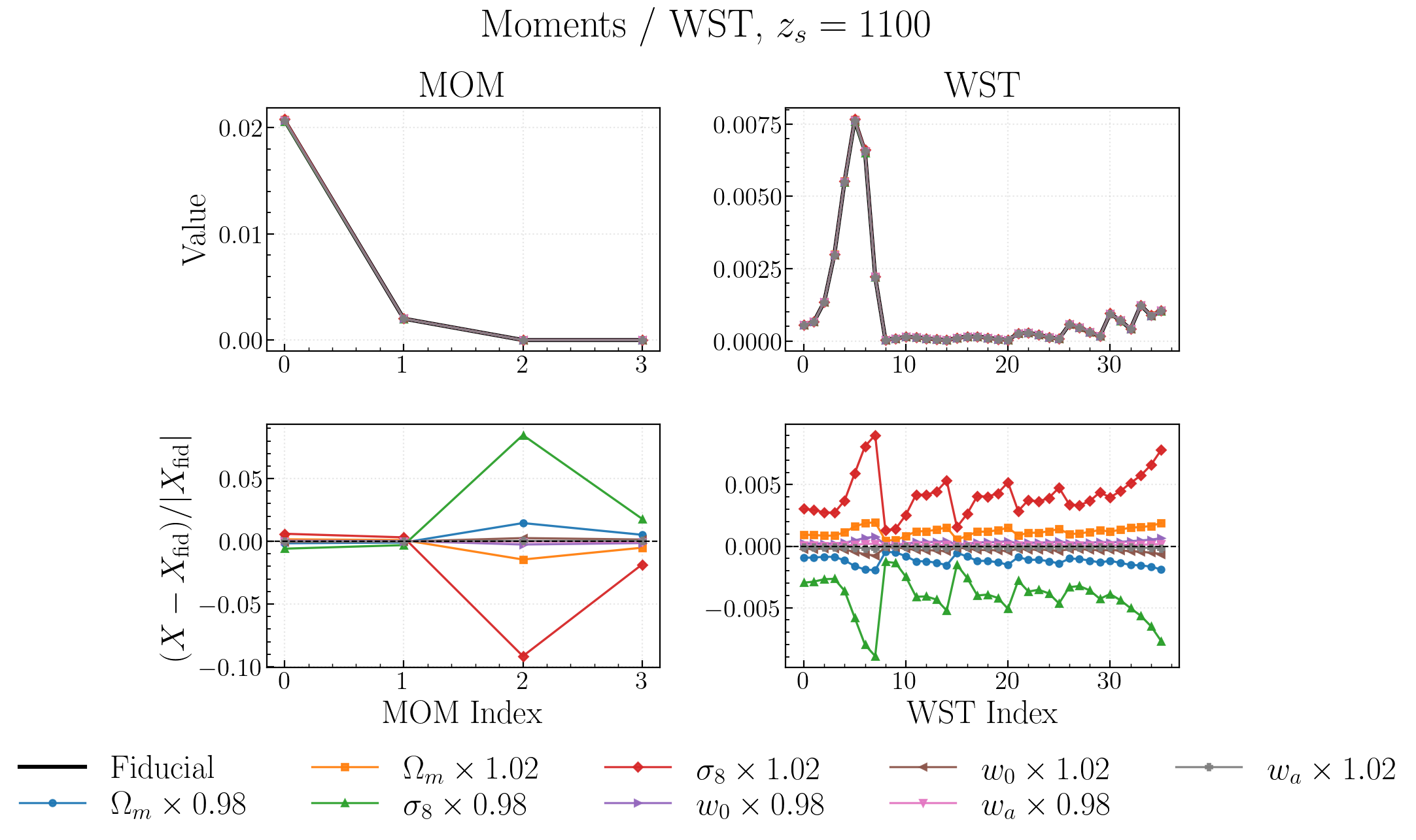}
    \caption{Same as Figure~\ref{fig:fractional_z10}, but for CMB lensing with source redshift $z_s=1100$.}
    \label{fig:fractional_z1100}
\end{figure*}

Figures~\ref{fig:fractional_z10} and~\ref{fig:fractional_z1100} show the statistics and their fractional parameter responses relative to the fiducial cosmology for two representative source redshifts, $z_s=1$ and $z_s=1100$, respectively. We consider shifts of $\pm2\%$ from the fiducial values for each parameter, holding the others fixed.

The low-redshift galaxy lensing statistics exhibit larger fractional parameter responses to all four parameters. In contrast, the non-Gaussian CMB lensing statistics respond only weakly to changes in $w_0$ and $w_a$. This behavior is consistent with the broad CMB lensing kernel, which receives substantial contributions from higher redshifts where dark energy is less dynamically important, thereby reducing its sensitivity to the late-time evolution of the dark energy equation of state. The relatively weak sensitivity of CMB lensing to dynamical dark energy has also been found in previous studies~\cite{Yu:2021vce,Planck:2015bue}.

\subsection{Likelihood}
We assume a Gaussian likelihood for all summary statistics. The covariance contains two contributions: the sample variance and measurement noise of the statistics, and the predictive uncertainty of the emulator.

We estimate the first contribution from 100,000 realizations at the fiducial $\Lambda$CDM cosmology:
\begin{align}
    \boldsymbol{C}_1 = \frac{1}{N-p-1}\sum_i(\boldsymbol{s}_i-\bar{\boldsymbol{s}})(\boldsymbol{s}_i-\bar{\boldsymbol{s}})^T\,,
\end{align}
where $\boldsymbol{s}_i$ is the data vector from the $i^{\rm th}$ simulation realization and $\bar{\boldsymbol{s}}$ is the corresponding sample mean. Because these statistics are measured on maps with an area of $12.25\,\mathrm{deg}^2$, we rescale this covariance by $r_{\rm sky}=12.25/15000$ to match the survey area. We also apply a Hartlap correction~\cite{Hartlap:2006kj}.

The second contribution, $\boldsymbol{C}_2$, accounts for the emulator's predictive uncertainty and has the analytical form given in Ref.~\cite{Chen:2026jlr}. The total covariance is therefore
\begin{align}
    \boldsymbol{C}_{\rm tot} = r_{\rm sky}\boldsymbol{C}_1+\boldsymbol{C}_2\,.
\end{align}
To improve the numerical stability of the covariance inversion, we normalize the data vector according to the related variance and regularize the covariance by adding a diagonal identity matrix multiplied by $10^{-3}$ to have a lower bound on the smallest eigenvalue. The correlation matrix for selected source redshifts and the convergence of the covariance estimate with the number of simulations used are shown in Appendix~\ref{appendix:correlation}. We also show the relative amplitude of $r_{\rm sky}\boldsymbol{C}_1$ and $\boldsymbol{C}_2$ in both Figure~\ref{fig:loocv_10} and~\ref{fig:loocv_11000} for $z_s=1$ and $1100$, respectively.


We impose independent Gaussian priors on the four parameters, $\Omega_m$, $\sigma_8$, $w_0$, and $w_a$. The prior widths are similar to (but about twice larger than) the constraints reported without the $3\times2$-point information in Table S1 of Ref.~\cite{DES:2026jmi}\footnote{The reported constraint of this case in Ref.~\cite{DES:2026jmi} is $(\sigma(\Omega_m), \sigma(\sigma_8), \sigma(w_0), \sigma(w_a))=(0.0055, 0.0095, 0.055, 0.21)$ and $(0.005, 0.008, 0.05, 0.195)$ when combining all data.}:
\begin{alignat}{2}
\label{eq.priors}
    &\sigma_{\rm prior}(\Omega_m)=0.01,\quad &\sigma_{\rm prior}(\sigma_8)&=0.02,\quad\nonumber\\
    &\sigma_{\rm prior}(w_0)=0.1,\quad &\sigma_{\rm prior}(w_a)&=0.3\,.
\end{alignat}
We adopt slightly broader widths than those reported in Ref.~\cite{DES:2026jmi}, as our prior does not account for correlations between parameters.

The forecast is then implemented with {\tt emcee}~\cite{foreman2013emcee}\footnote{\url{https://emcee.readthedocs.io/en/stable/}}, an affine invariant Markov chain Monte Carlo (MCMC) ensemble sampler. We run 32 parallel chains and ensure that the length of the chain is at least 100 times longer than the auto-correlation time~\cite{2010CAMCS...5...65G}. Note again that we set the fiducial cosmology for sampling to be at $(\Omega_m,\sigma_8,w_0,w_a)=(0.315, 0.82, -0.75, -0.75)$, which is different from the cosmology at which the covariance matrix is evaluated at. 

\section{Validation and Null Tests}\label{sec:validation_and_null_tests}
We perform several validation and null tests to assess the reliability of the simulation pipeline and the emulators. First, we test the emulator predictions at the fiducial cosmology after excluding it from the training set. We then compare the power-spectrum constraints obtained from the simulation-based emulator with those derived from analytical predictions. Finally, we repeat the inference using Gaussian random fields to test whether adding non-Gaussian statistics produces a spurious improvement over the power-spectrum constraints. These tests assess the accuracy of the emulator predictions, the consistency of the forecast pipeline, and the interpretation of the additional constraining power.

\subsection{Leave-One-Out Cross-Validation}
We validate the emulator for each statistic by withholding the fiducial cosmology from the training set. We train on the remaining cosmologies and use the resulting emulator to predict the statistics at the excluded cosmology. Comparing these predictions with the directly measured statistics from the fiducial-cosmology simulation tests the emulator at the point around which our forecasts are constructed. We evaluate the prediction errors relative to the expected survey statistical uncertainties to assess their relevance for the parameter forecasts.

\begin{figure*}
    \centering
    \includegraphics[width=0.24\linewidth]{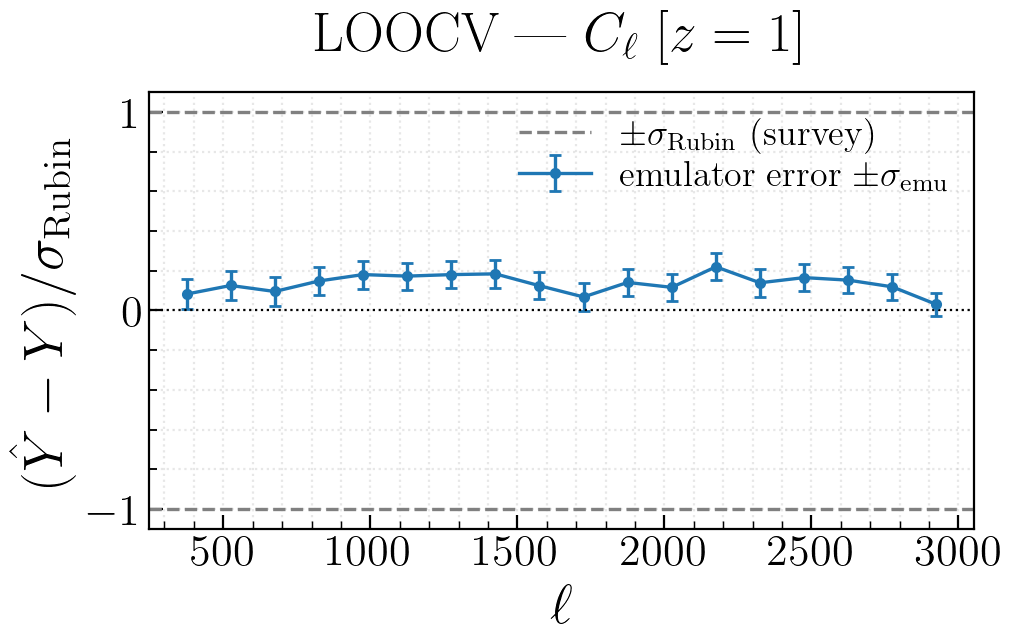}
    \includegraphics[width=0.24\linewidth]{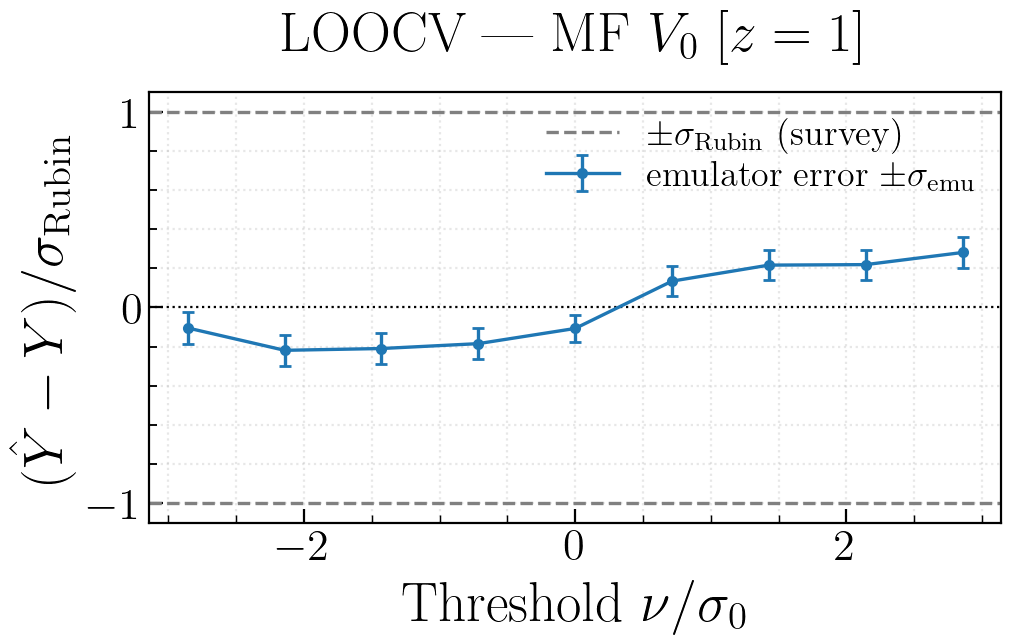}
    \includegraphics[width=0.24\linewidth]{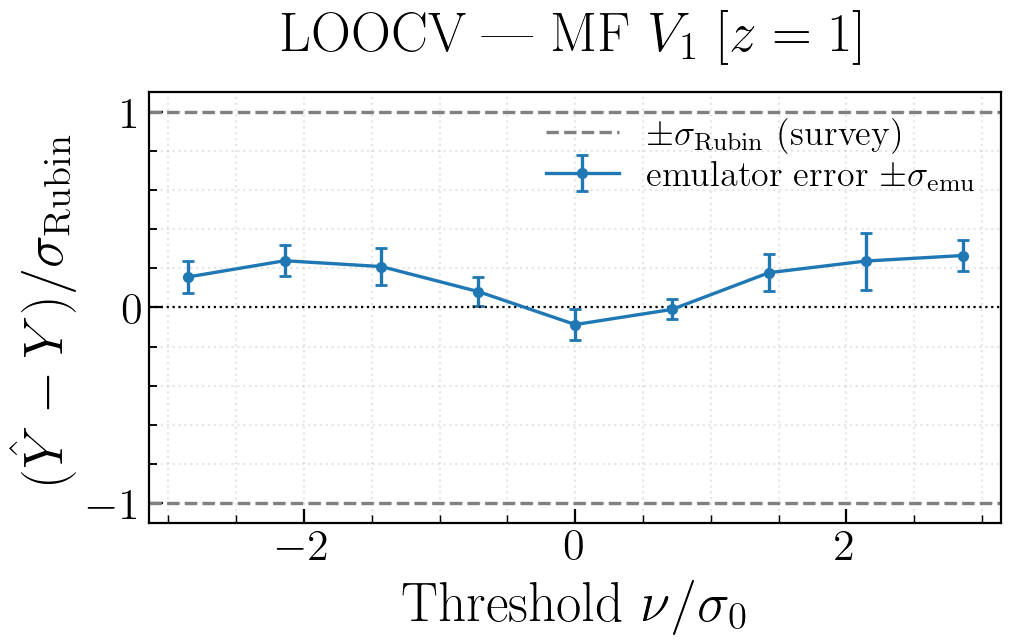}
    \includegraphics[width=0.24\linewidth]{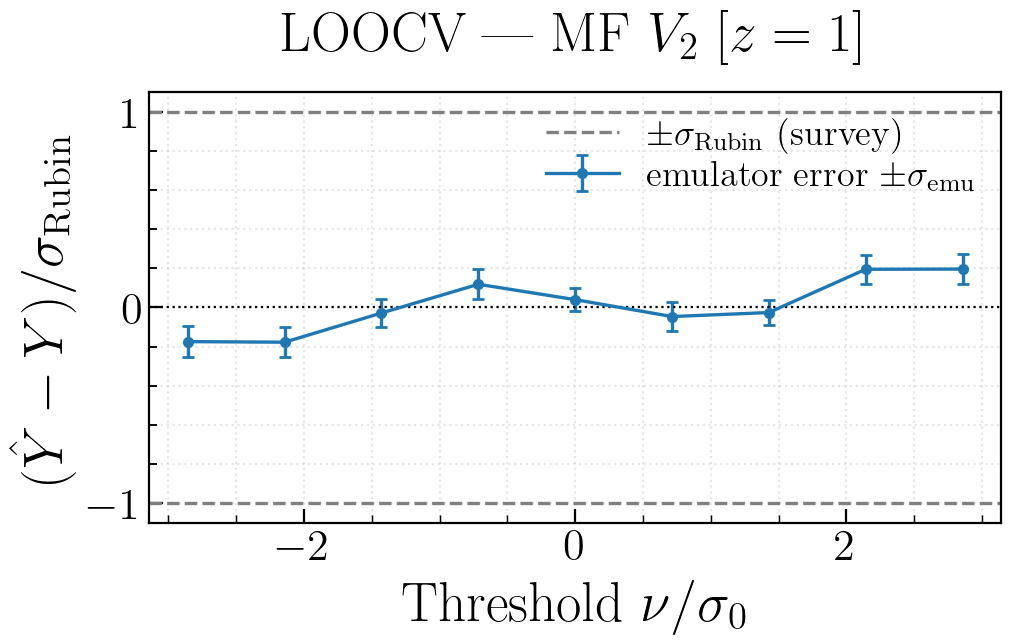}
    \includegraphics[width=0.24\linewidth]{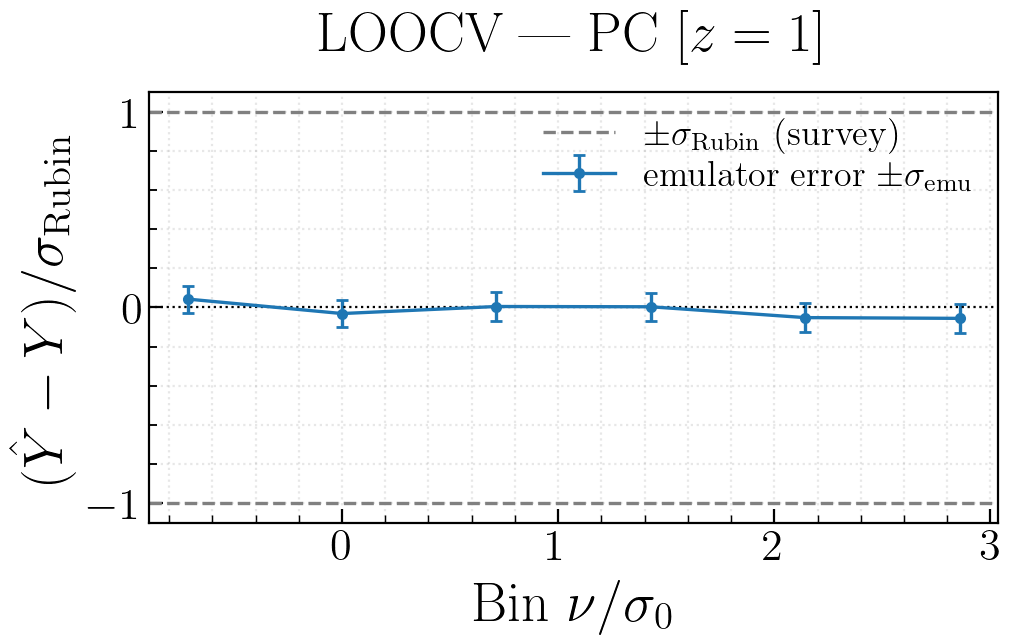}
    \includegraphics[width=0.24\linewidth]{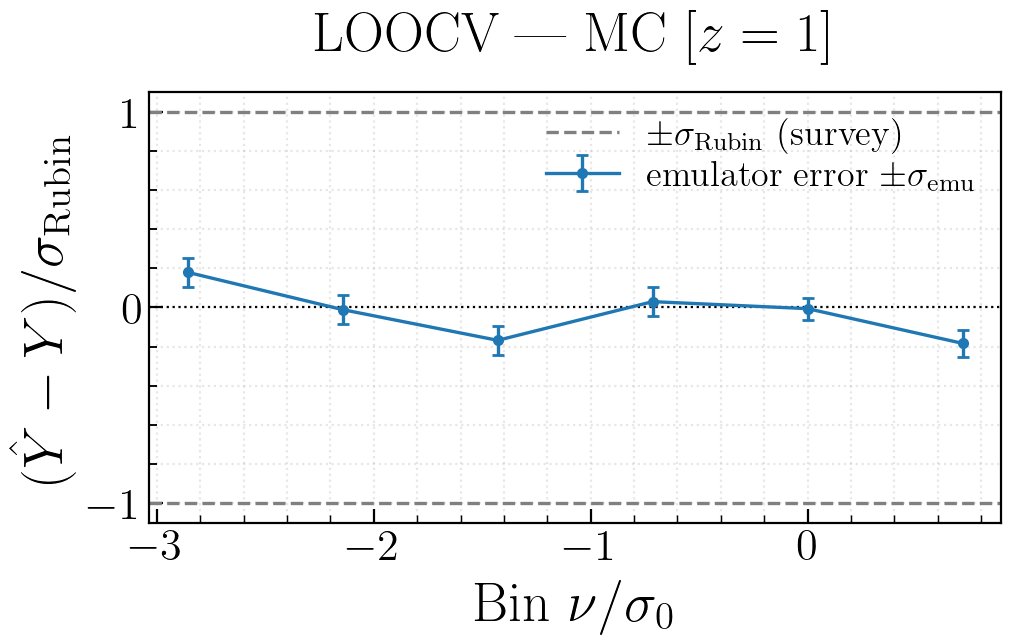}
    \includegraphics[width=0.24\linewidth]{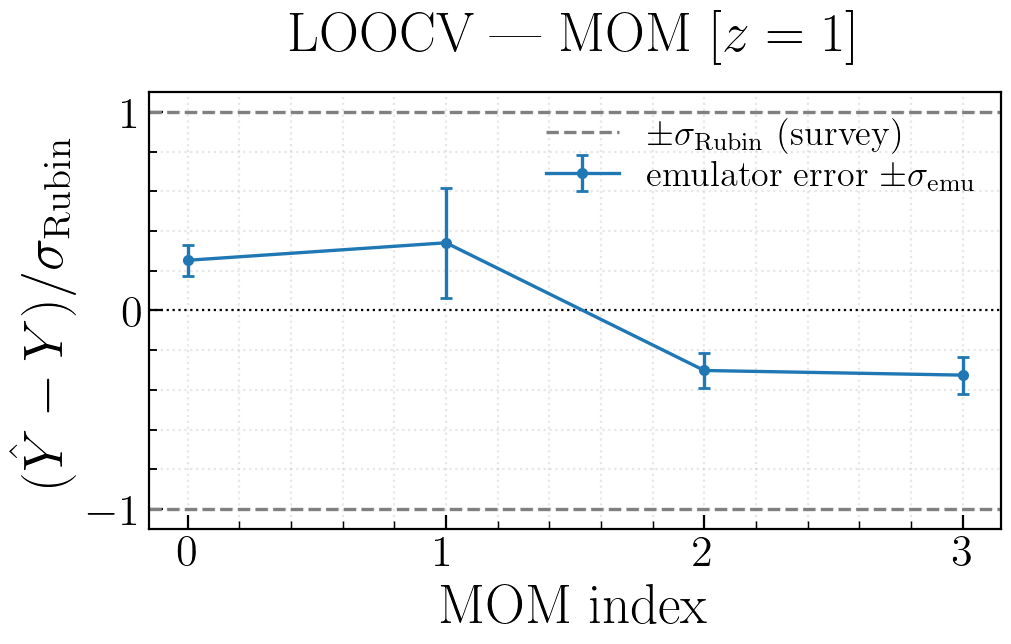}
    \includegraphics[width=0.24\linewidth]{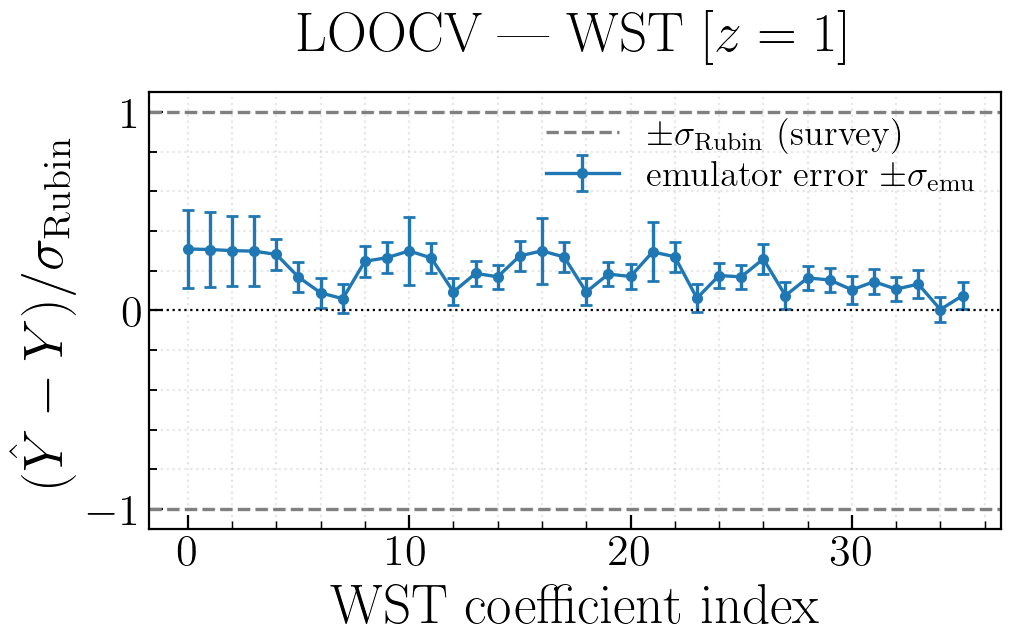}
    \caption{Leave-one-out validation of the Gaussian process emulator at source redshift $z_s=1$. The emulator is trained on all sampled cosmologies except the fiducial cosmology and used to predict the fiducial data vector. The blue curve shows the fractional prediction error $(\hat{Y}-Y)/\sigma_{\rm Rubin}$, with the error bar indicating the emulator uncertainty when the full data vector is included. From top left to bottom right, the panels show the power spectrum $C_\ell$ over $300\leq\ell\leq3000$, the Minkowski functionals $V_0$, $V_1$, and $V_2$, peak counts (PC), minima counts (MC), moments (MOM), and the wavelet scattering transform (WST), all measured from Wiener-filtered convergence maps. MFs, PC, and MC are shown as functions of normalized threshold or bin amplitude, $\nu/\sigma_0$; MOM and WST coefficients are labeled by index. The grey dashed lines show the $\pm1\sigma$ statistical uncertainty for a $15,000\,\mathrm{deg}^2$ Rubin-like survey, obtained by rescaling the covariance estimated from the $(3.5\,\mathrm{deg})^2$ map realizations. The prediction errors remain within the survey statistical uncertainties for every statistic.}
    \label{fig:loocv_10}
\end{figure*}

\begin{figure*}
    \centering
    \includegraphics[width=0.24\linewidth]{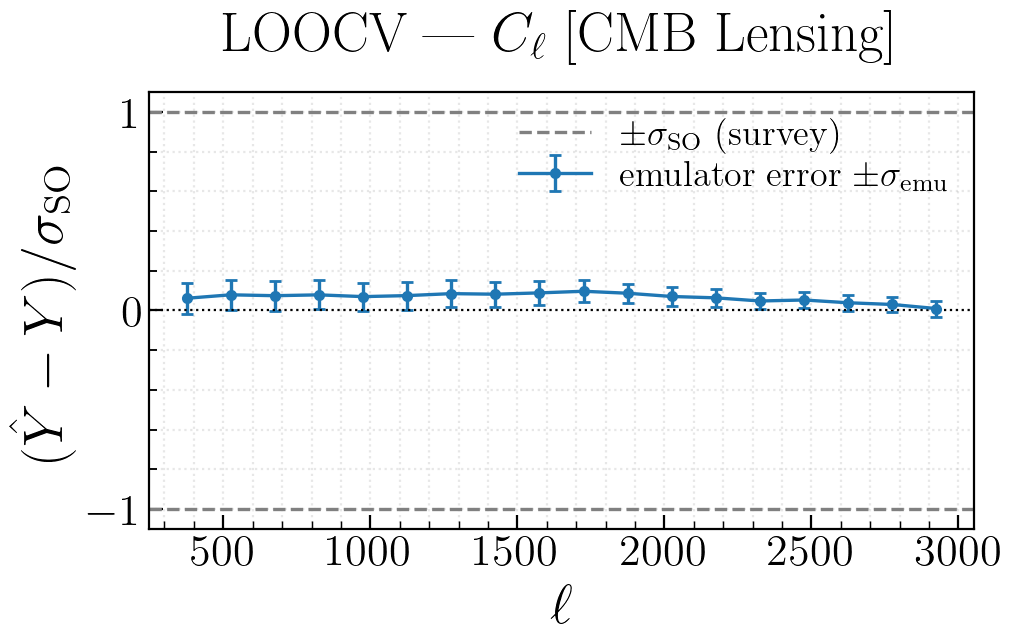}
    \includegraphics[width=0.24\linewidth]{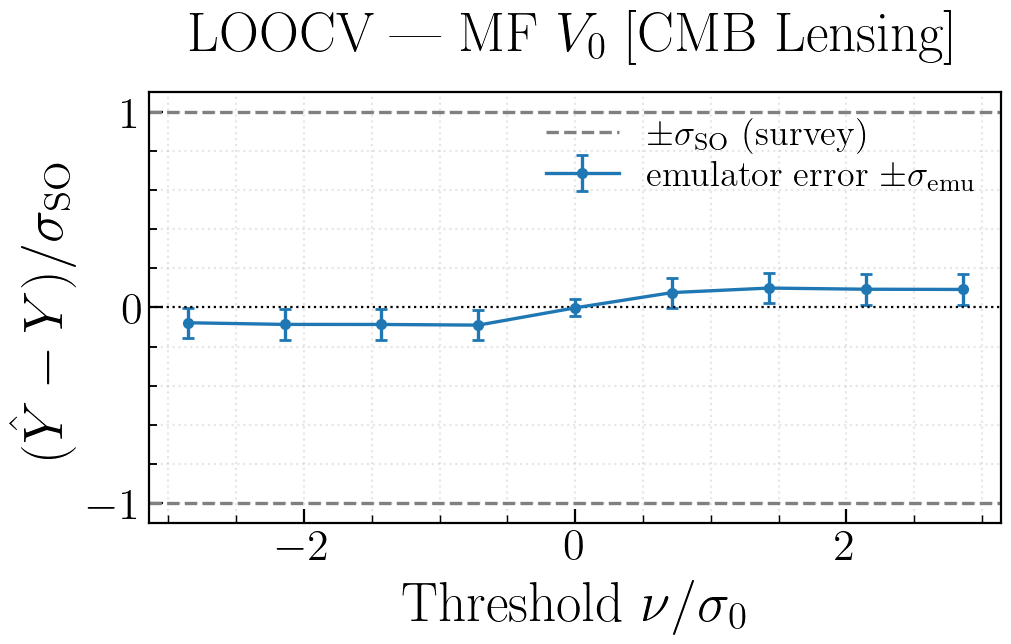}
    \includegraphics[width=0.24\linewidth]{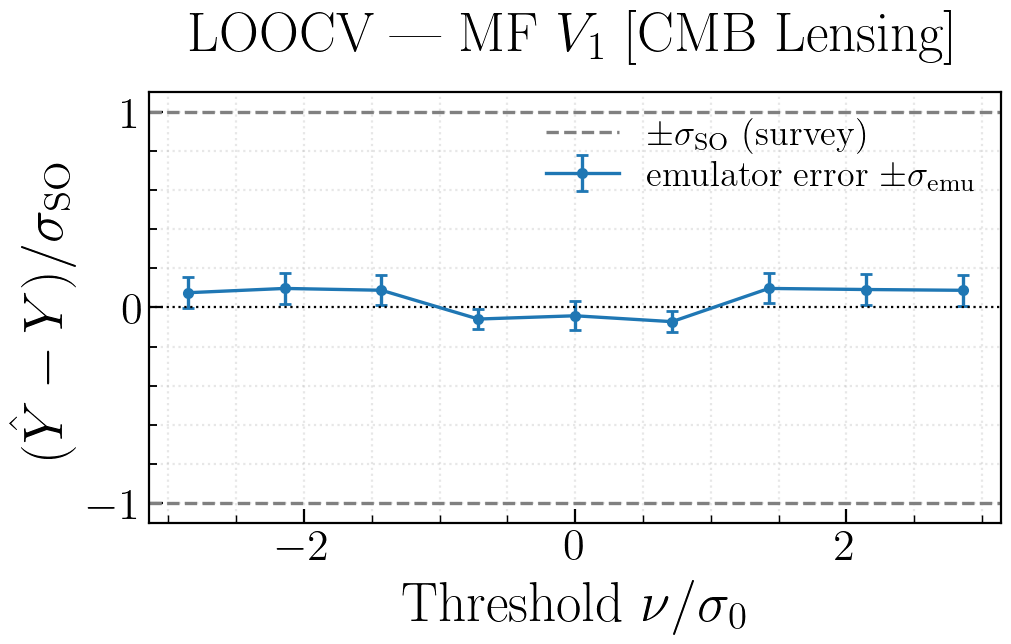}
    \includegraphics[width=0.24\linewidth]{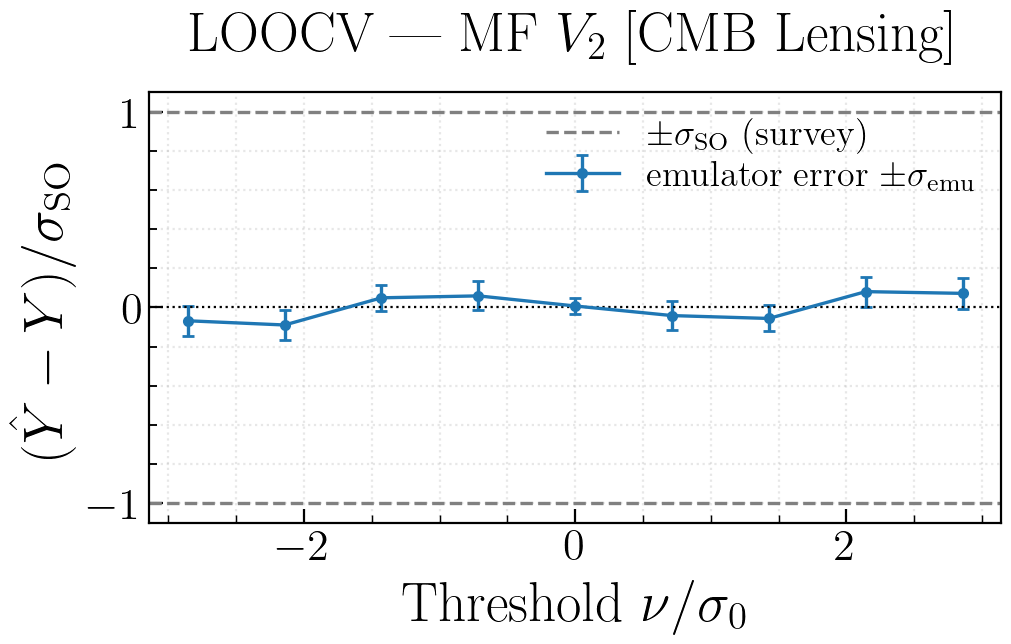}
    \includegraphics[width=0.24\linewidth]{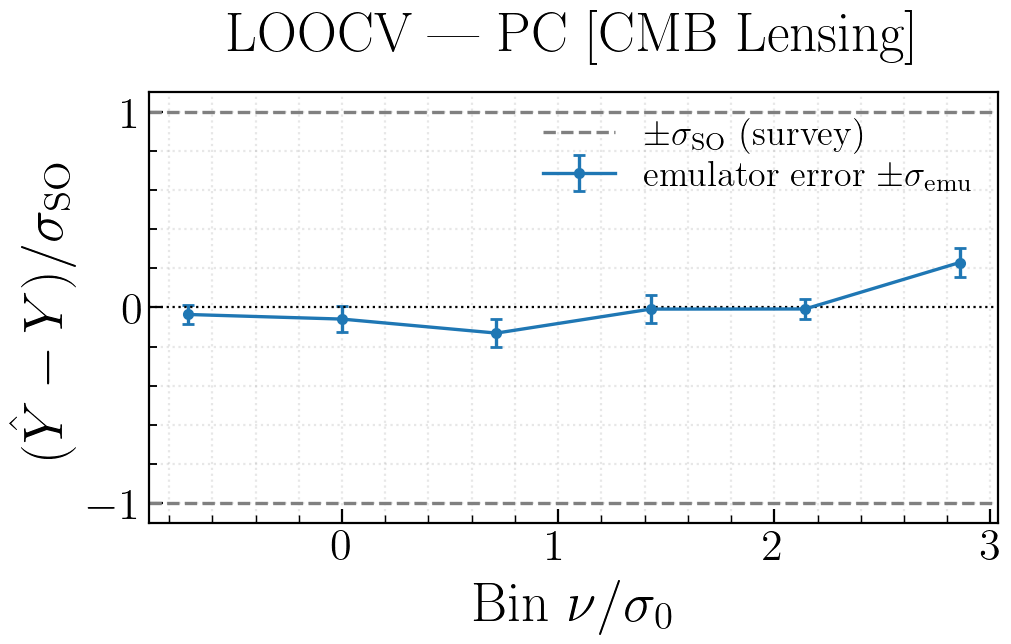}
    \includegraphics[width=0.24\linewidth]{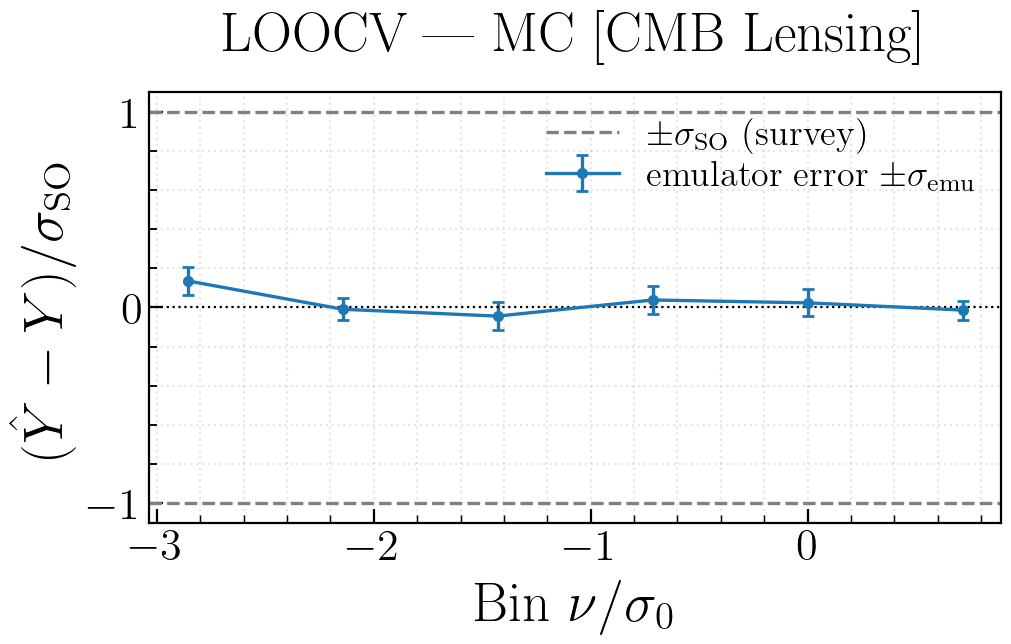}
    \includegraphics[width=0.24\linewidth]{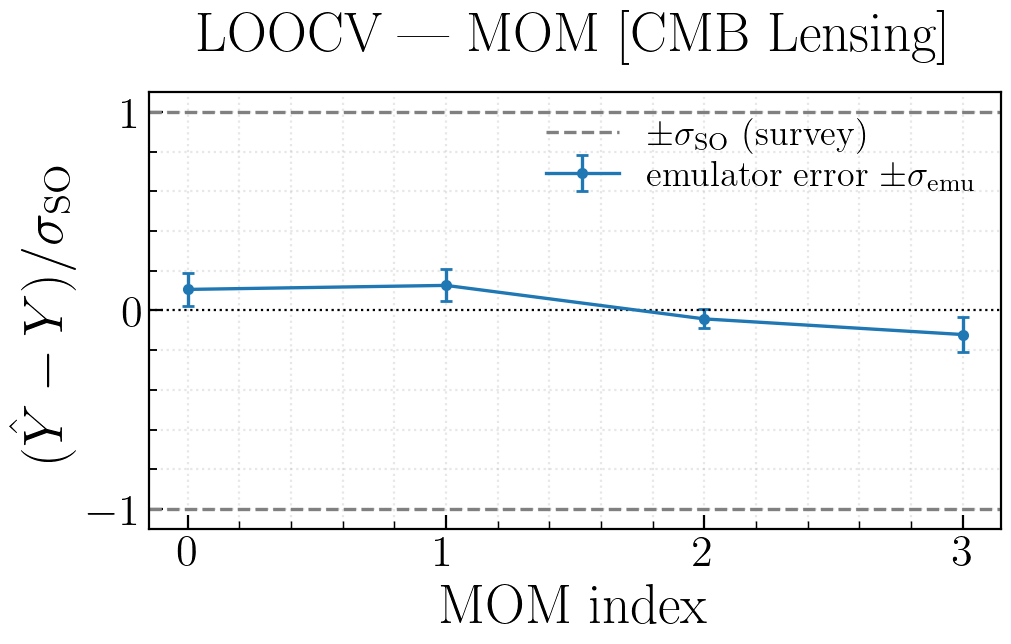}
    \includegraphics[width=0.24\linewidth]{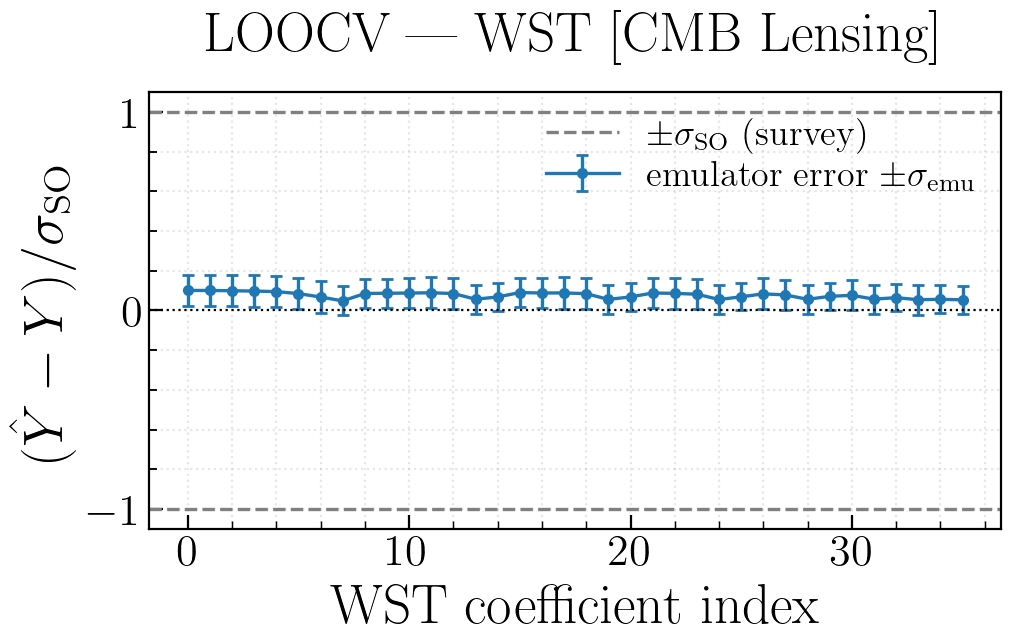}
    \caption{Same as Figure~\ref{fig:loocv_10}, but for CMB lensing with source redshift $z_s=1100$ and SO-like noise.}
    \label{fig:loocv_11000}
\end{figure*}

Figures~\ref{fig:loocv_10} and~\ref{fig:loocv_11000} show this comparison for two representative source redshifts, using the statistical uncertainties expected for $15,000\,\mathrm{deg}^2$ Rubin-like and SO-like surveys. In both cases, the prediction errors lie within the corresponding survey statistical uncertainties for all the statistics considered with at most an error of $0.4\sigma_{\rm Rubin}$. This comparison provides a check of the emulator predictions at the fiducial cosmology. We additionally account for the emulator's predictive uncertainty through the covariance contribution $\boldsymbol{C}_2$ in the likelihood.

\subsection{Validation against Analytical Methods}
We validate the simulations and the power-spectrum emulator by comparing their parameter forecasts with those obtained from analytical power spectra computed using {\tt CAMB}~\cite{Lewis:1999bs}\footnote{\url{http://camb.info}} and {\tt PYCCL}. For the nonlinear matter power spectrum, we adopt {\tt HMcode-2020}~\cite{Mead:2020vgs}. Although its fitting parameters were calibrated using simulations with $w_a=0$, HMcode can still be applied to dynamical dark energy cosmologies because the dependence on $w_a$ still enters through the background expansion and linear growth. The model was tested against the Mira-Titan simulations~\cite{Lawrence:2017ost,Bocquet:2020tes}, which include variations in both $w_0$ and $w_a$, and achieves a root-mean-squared error of approximately $2.5\%$, although the error can reach $\sim16\%$ for the most extreme cosmologies. For the simulation-based emulator, we obtain the constraints through MCMC sampling, whereas we use a Fisher forecast for the analytical predictions. We adopt the same data covariance in both calculations, allowing us to compare the resulting parameter uncertainties and degeneracy directions for a common experimental configuration.

\begin{figure}
    \centering
    \includegraphics[width=1.0\linewidth]{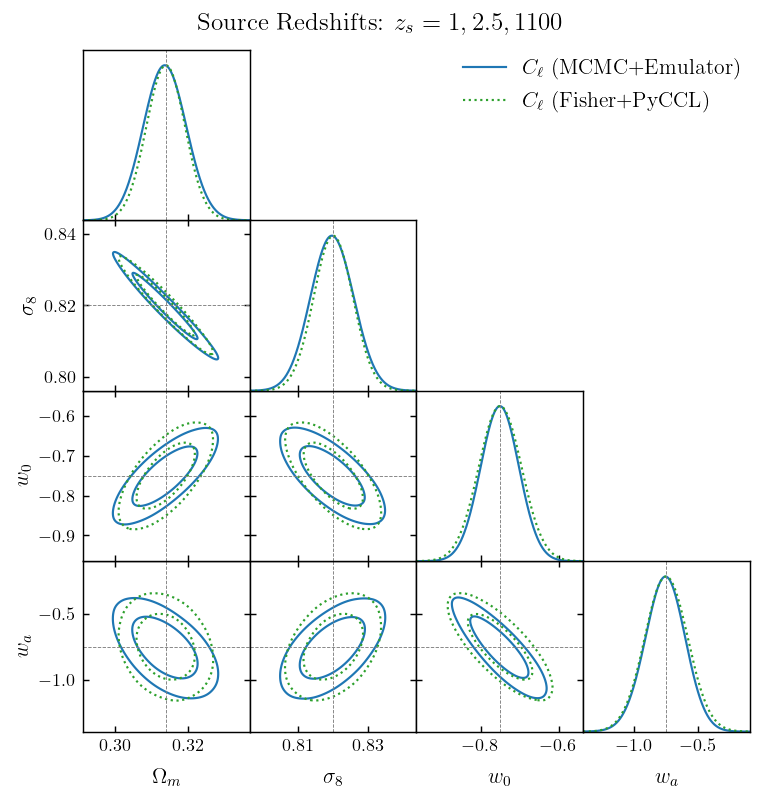}
    \caption{Comparison of power-spectrum-only cosmological constraints obtained from the simulation-based emulator (blue, solid) and analytical predictions from {\tt PyCCL} and {\tt CAMB} (green, dotted). We use MCMC sampling for the emulator constraints and a Fisher forecast for the analytical predictions, adopting the same data covariance in both cases. Contours show the marginalized $68\%$ and $95\%$ regions, and dashed lines mark the fiducial parameter values. We analyze auto- and cross-power spectra for source redshifts $z_s=[1,2.5,1100]$ and use $\ell\in[300,1500]$. The marginalized $1\sigma$ uncertainties agree to within approximately $10\%$ for all parameters.}
    \label{fig:validation_ccl_test_triangleplot}
\end{figure}

Figure~\ref{fig:validation_ccl_test_triangleplot} shows this comparison for the source-redshift combination $z_s=[1,2.5,1100]$, using multipoles in the range $\ell\in[300,1500]$. The marginalized uncertainties agree to within approximately $10\%$ for all four parameters. The agreement improves when higher-redshift sources are included, or when only higher-redshift sources are used, where nonlinear evolution is less pronounced. We note that the discrepancy is largest in the $w_a$ panels, in agreement with Figure~7 of Ref.~\cite{Chen:2026zqq}, which shows that the differences become more significant when moving beyond $\Lambda$CDM, and we have $w_0=w_a=-0.75$ in our fiducial model here.
This comparison provides a complementary check of the power-spectrum forecasts in addition to the validation of the emulator predictions at the fiducial cosmology.

\subsection{Null Tests with Gaussian Random Field}
We perform a Gaussian null test to check whether the non-Gaussian-statistic emulators spuriously infer information beyond the power spectrum. For each sampled cosmology, we measure the power spectra of the lensing convergence maps and use them to generate Gaussian random fields. We generate 10,000 realizations for each cosmology and source redshift.

\begin{figure*}
    \centering
    \includegraphics[width=1.0\linewidth]{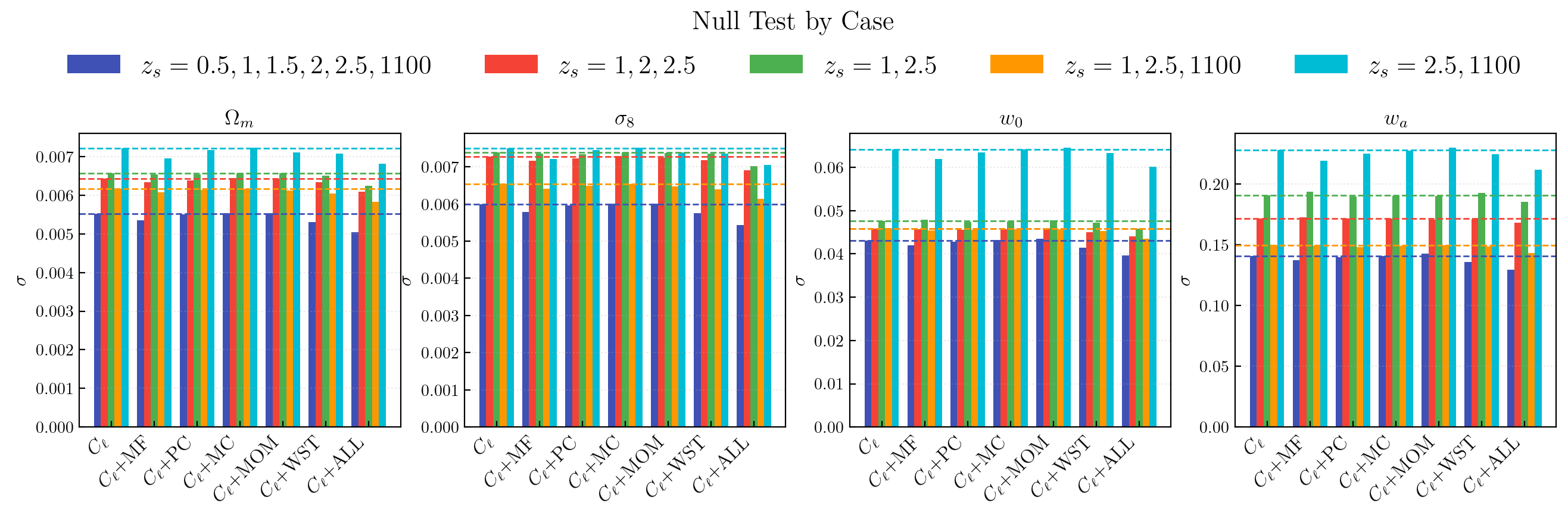}
    \caption{Marginalized $1\sigma$ uncertainties on $\Omega_m$, $\sigma_8$, $w_0$, and $w_a$ (left to right) in the Gaussian null test. Within each panel, the groups show constraints from $C_\ell$ alone; $C_\ell$ combined with Minkowski functionals (MF), peak counts (PC), minima counts (MC), moments (MOM), the wavelet scattering transform (WST), or all five jointly ($C_\ell+\mathrm{ALL}$). Bar colors indicate the source-redshift combinations, with $z_s=1100$ denoting CMB lensing, and dashed lines indicate the constraints from $C_\ell$ only. Adding non-Gaussian statistics to $C_\ell$ yields no appreciable improvement (as expected for Gaussian random field maps). These results support the consistency of the emulation and inference pipeline.}
    \label{fig:null_test_barplot}
\end{figure*}

\begin{figure}
    \centering
    \includegraphics[width=1.0\linewidth]{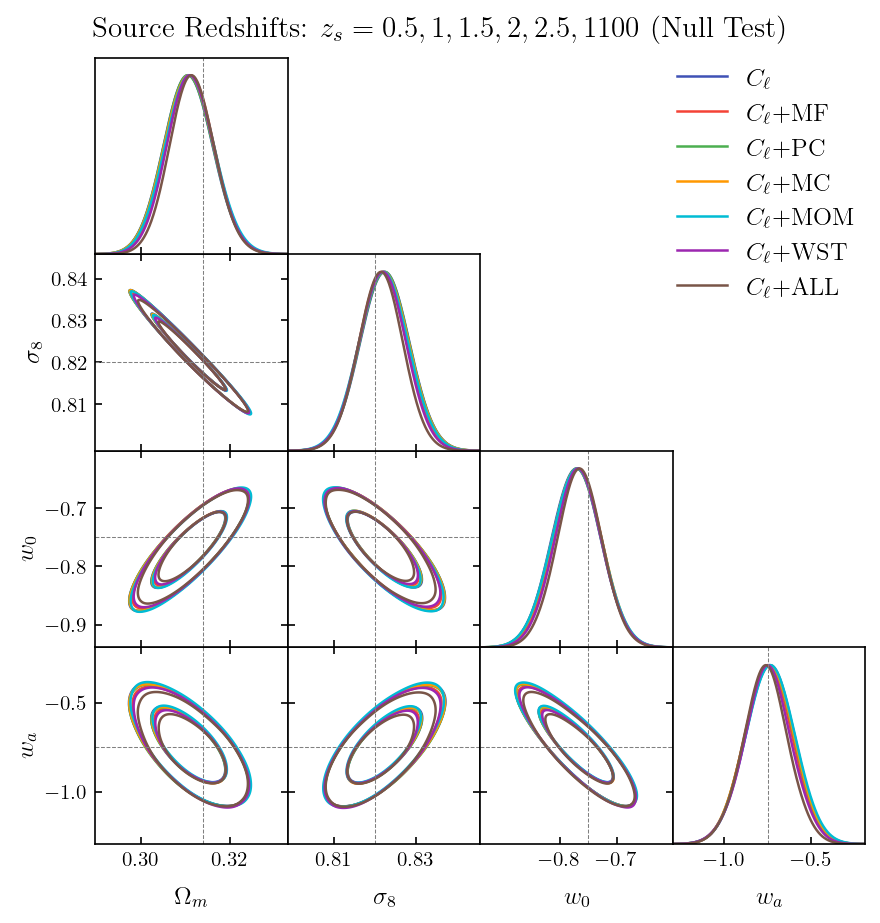}
    \caption{Marginalized $68\%$ and $95\%$ posterior contours for the Gaussian-map null test in Figure~\ref{fig:null_test_barplot}, using the source-redshift combination $z_s=[0.5,1,1.5,2,2.5,1100]$. The parameter degeneracy directions agree closely across the summary-statistic combinations, illustrating that the non-Gaussian statistics provide no additional information in this analysis (as expected for Gaussian random field maps).}
    \label{fig:null_test_triangleplot}
\end{figure}

Figure~\ref{fig:null_test_barplot} compares the marginalized uncertainties for different combinations of source redshifts and summary statistics. Adding any of the non-Gaussian statistics to $C_\ell$ yields no appreciable improvement in the parameter constraints, supporting the consistency of the emulation and inference pipeline. Figure~\ref{fig:null_test_triangleplot} shows the posterior contours for the combination $z_s=[0.5,1,1.5,2,2.5,1100]$, which has the highest number of source redshifts we consider. The parameter degeneracy directions also agree closely across the different summary-statistic combinations.

\section{Results}\label{sec:results}

\subsection{Improvement from non-Gaussian statistics}

\begin{figure*}
    \centering
    \includegraphics[width=0.45\linewidth]{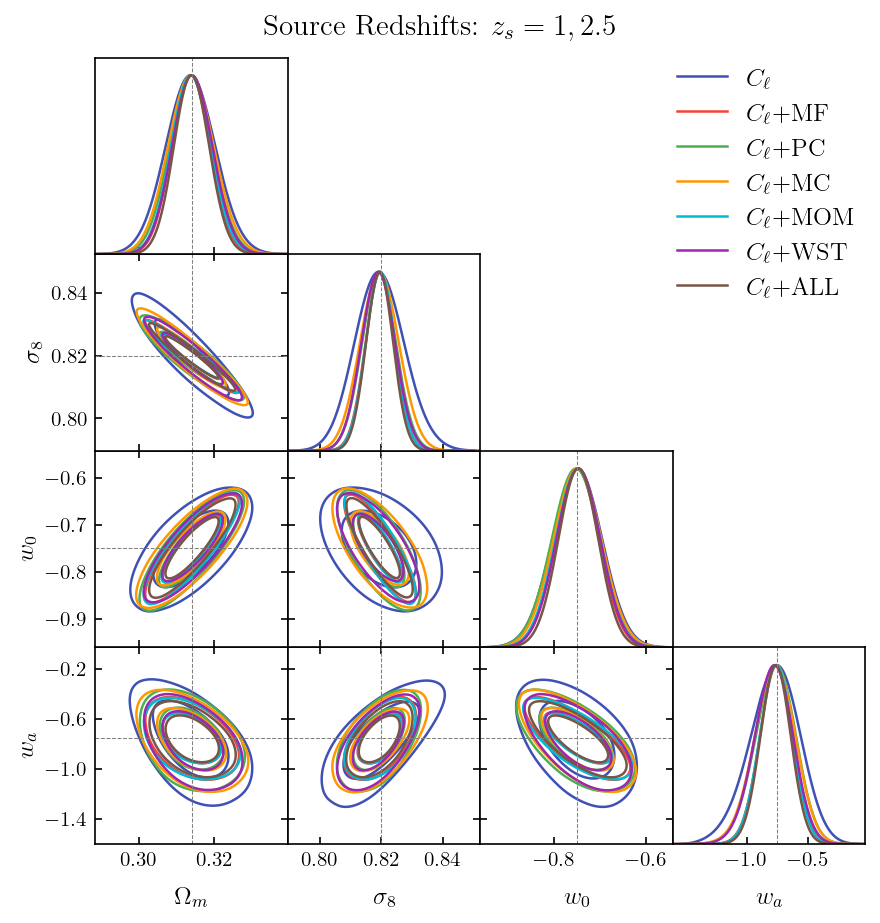}
    \includegraphics[width=0.45\linewidth]{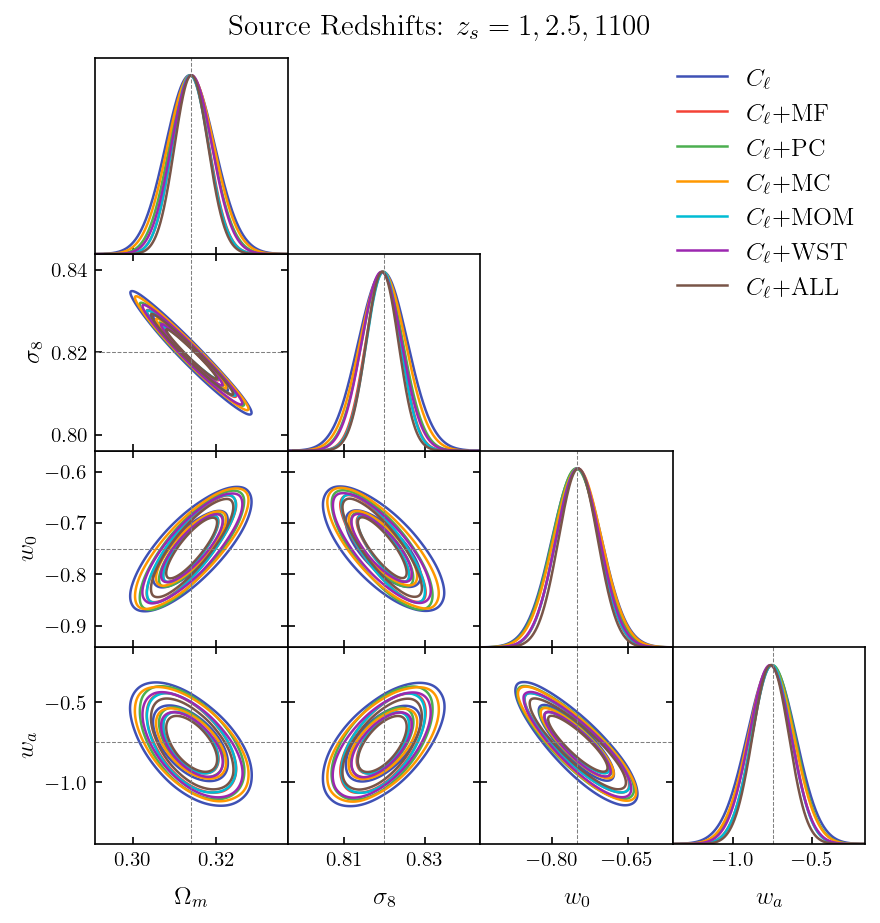}
    \includegraphics[width=0.45\linewidth]{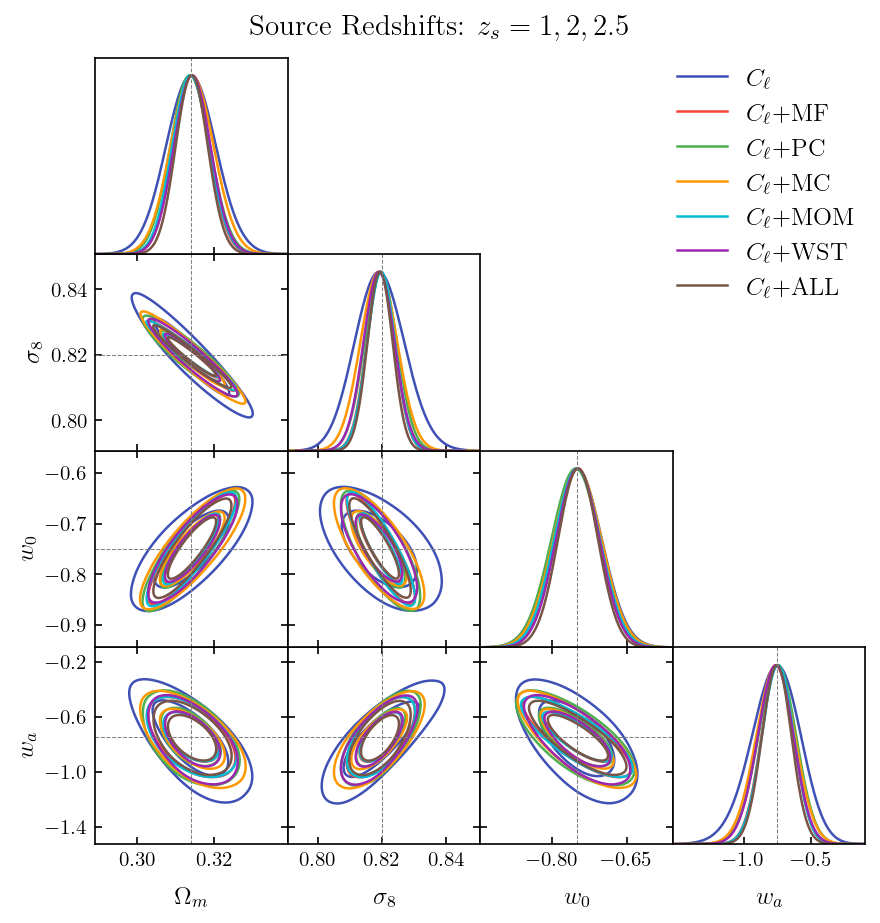}
    \includegraphics[width=0.45\linewidth]{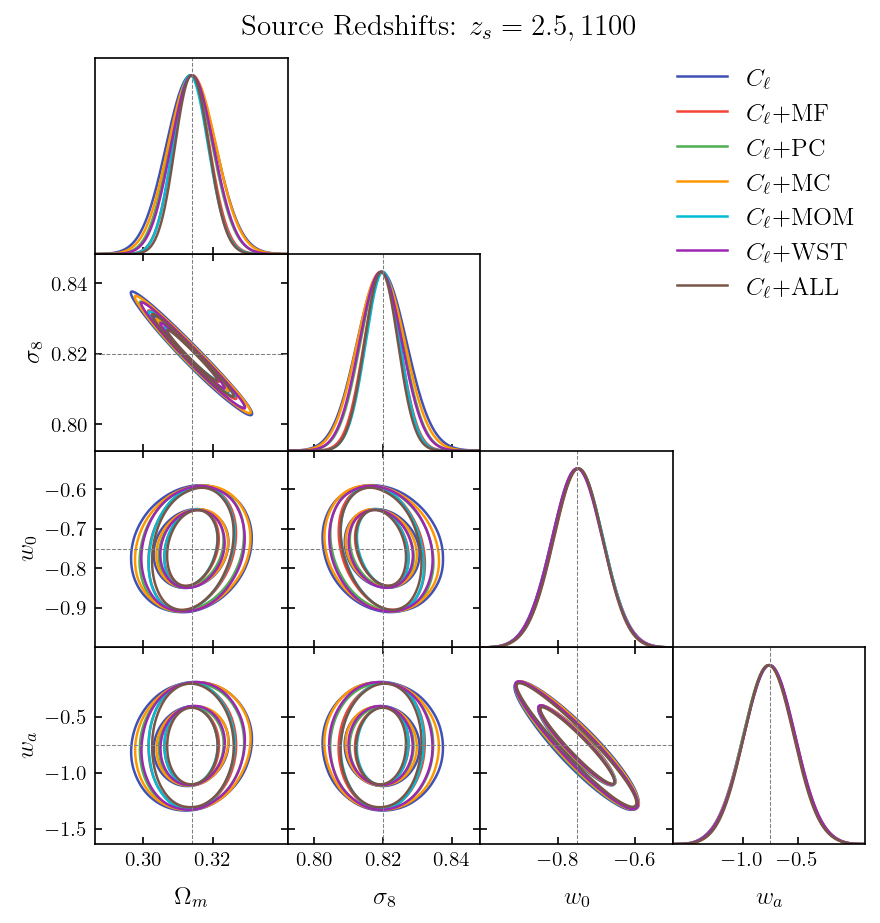}
    \caption{Cosmological constraints obtained by combining non-Gaussian statistics with the tomographic lensing auto- and cross-power spectra for four source-redshift combinations, as labeled in the plot titles. All statistics are measured from Wiener-filtered convergence maps, and $z_s=1100$ denotes CMB lensing. Contours show the marginalized $68\%$ and $95\%$ credible regions for $(\Omega_m,\sigma_8,w_0,w_a)$, with all four parameters varied. Dashed lines mark the fiducial values. In each panel, the baseline constraints (blue) use only the auto- and cross-spectra $C_\ell$. The other contours add Minkowski functionals (MF), peak counts (PC), minima counts (MC), moments (MOM), the wavelet scattering transform (WST), or all five jointly ($C_\ell+\mathrm{ALL}$).}
    \label{fig:forecast_triangle}
\end{figure*}

\begin{figure}
    \centering
    \includegraphics[width=1.0\linewidth]{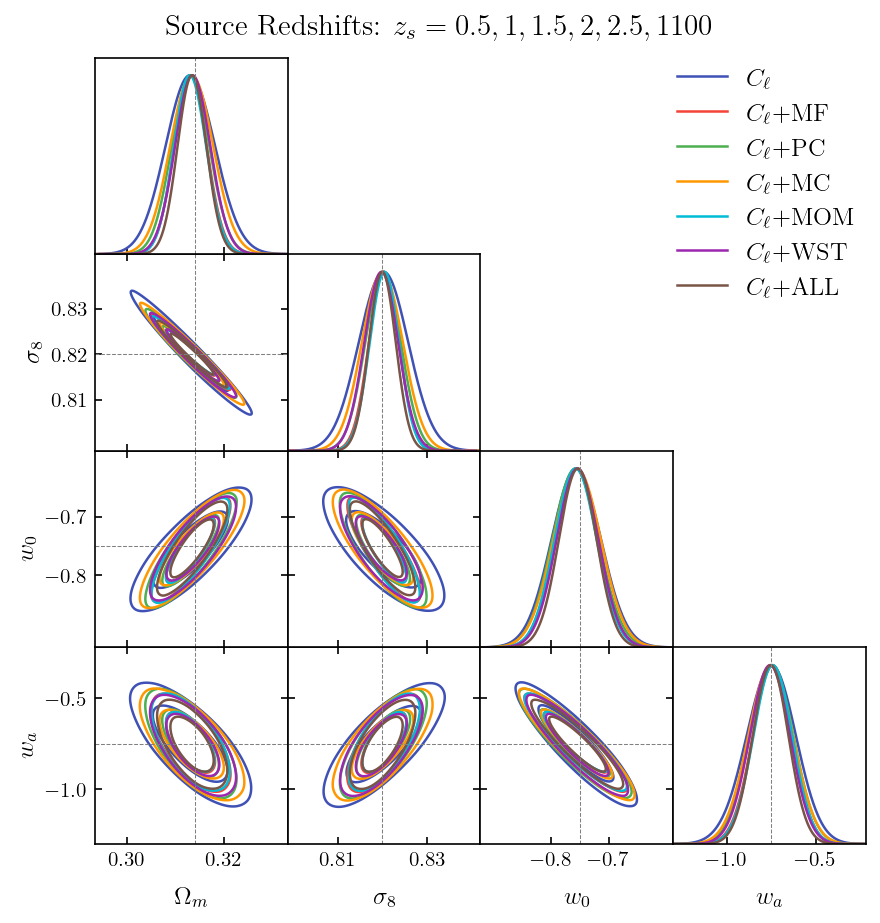}
    \caption{Cosmological constraints obtained by combining non-Gaussian statistics with the tomographic lensing power spectra using all available source redshifts. All statistics are measured from Wiener-filtered convergence maps, and $z_s=1100$ denotes CMB lensing. Contours show the marginalized $68\%$ and $95\%$ credible regions for $(\Omega_m,\sigma_8,w_0,w_a)$, with all four parameters varied. Dashed lines mark the fiducial values. In each panel, the baseline constraints (blue) use only the auto- and cross-spectra $C_\ell$. The other contours add Minkowski functionals (MF), peak counts (PC), minima counts (MC), moments (MOM), the wavelet scattering transform (WST), or all five jointly ($C_\ell+\mathrm{ALL}$).}
    \label{fig:forecast_triangle_all}
\end{figure}

\begin{figure*}
    \centering
    \includegraphics[width=1.0\linewidth]{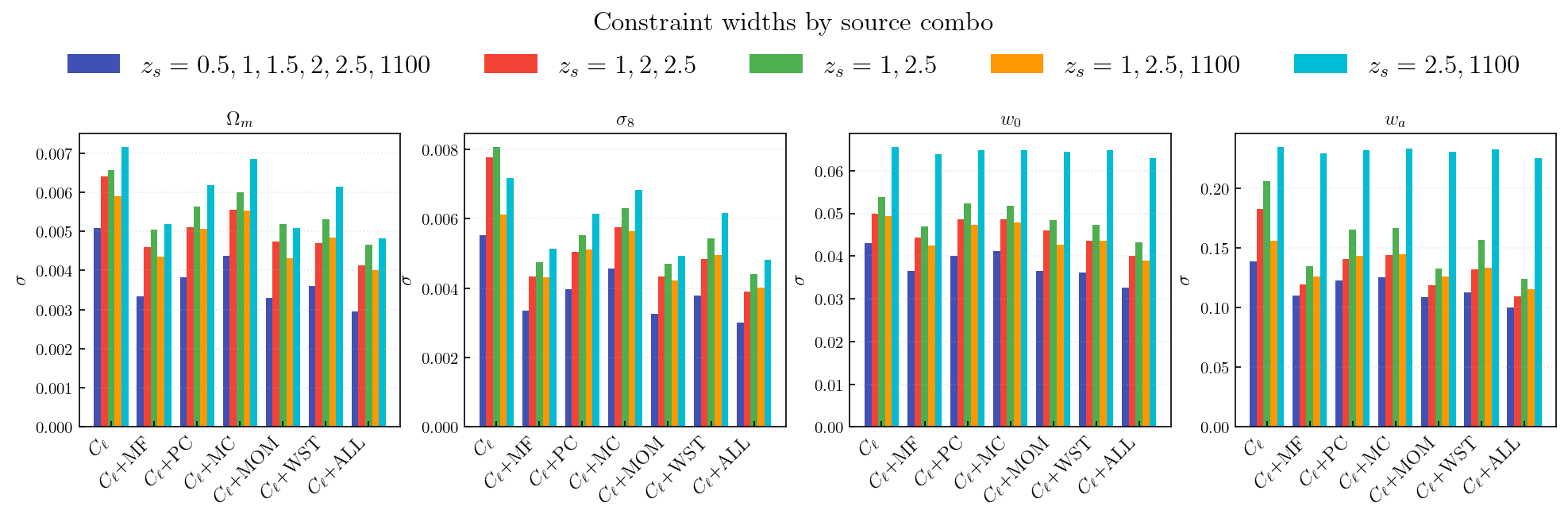}
    \caption{Marginalized $1\sigma$ uncertainties on $\Omega_m$, $\sigma_8$, $w_0$, and $w_a$ (left to right) for each data vector and source-redshift combination. Within each panel, the groups show constraints from the lensing power spectra alone ($C_\ell$) and from $C_\ell$ combined with Minkowski functionals (MF), peak counts (PC), minima counts (MC), moments (MOM), the wavelet scattering transform (WST), or all five jointly ($C_\ell+\mathrm{ALL}$). Bar colors indicate the source-redshift combinations, with $z_s=1100$ denoting CMB lensing.}
    \label{fig:forecast_barplot}
\end{figure*}

Figures~\ref{fig:forecast_triangle} and~\ref{fig:forecast_triangle_all} show the marginalized two-dimensional posterior contours for four representative source-redshift combinations and for the combination of all available source redshifts. Figure~\ref{fig:forecast_barplot} summarizes the corresponding marginalized one-dimensional uncertainties. We consider five configurations: $z_s=[1,2,2.5]$, $[1,2.5]$, $[1,2.5,1100]$, $[2.5,1100]$, and $[0.5,1,1.5,2,2.5,1100]$, where $z_s=1100$ denotes CMB lensing. Note that in our analysis, we do not consider a single $z_s$ as it will leave the dynamical dark energy unconstrained.

The results show that galaxy-lensing tomography is particularly important for constraining $w_0$ and $w_a$. For $z_s=[2.5,1100]$, which contains only one galaxy-lensing source plane together with CMB lensing, the constraints on the dark-energy parameters remain weak. Adding non-Gaussian statistics to the data vector in this case improves $\sigma(w_0)$ and $\sigma(w_a)$ by only a few percent, although the constraints on $\Omega_m$ and $\sigma_8$ improve by more than $30\%$ when all non-Gaussian statistics are combined. However, once multiple galaxy source planes are included, the redshift dependence of the lensing signal helps break degeneracies involving $w_0$ and $w_a$, and the non-Gaussian statistics provide substantial additional information on all four parameters.

\begin{figure*}
    \centering
    \includegraphics[width=1.0\linewidth]{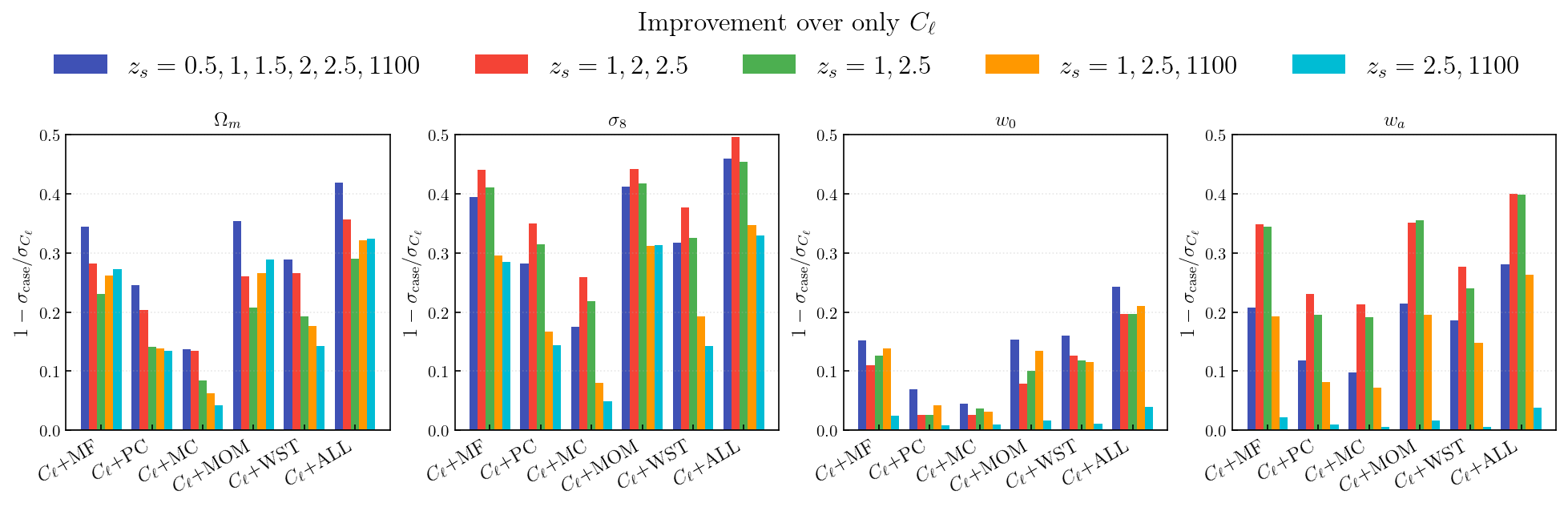}
    \caption{Fractional reduction in marginalized $1\sigma$ uncertainties relative to the power-spectrum-only constraints, $1-\sigma_{\rm case}/\sigma_{C_\ell}$. Positive values indicate tighter constraints. Panels show $\Omega_m$, $\sigma_8$, $w_0$, and $w_a$ from left to right. Within each panel, the groups correspond to the statistics added to $C_\ell$: Minkowski functionals (MF), peak counts (PC), minima counts (MC), moments (MOM), the wavelet scattering transform (WST), and all five jointly ($C_\ell+\mathrm{ALL}$). Bar colors indicate the source-redshift combinations, with $z_s=1100$ denoting CMB lensing.}
    \label{fig:forecast_improveCell_barplot}
\end{figure*}

Figure~\ref{fig:forecast_improveCell_barplot} shows the fractional reduction in marginalized uncertainties relative to the power-spectrum-only constraints. For $z_s=[1,2,2.5]$, combining all five non-Gaussian statistics reduces the uncertainties on $\Omega_m$, $\sigma_8$, $w_0$, and $w_a$ by $35.6\%$, $49.6\%$, $19.7\%$, and $40.0\%$, respectively. Similar improvements are obtained for $z_s=[1,2.5]$, with reductions of $28.9\%$, $45.4\%$, $19.7\%$, and $39.8\%$. When all available source redshifts are included ($z_s = [0.5,1,1.5,2,2.5,1100]$), the corresponding improvements are $41.9\%$, $45.8\%$, $24.3\%$, and $28.0\%$. Non-Gaussian statistics therefore continue to provide substantial additional cosmological information even after the full set of tomographic power spectra are included.

Table~\ref{tab:pk_improvement_allzs} compares the individual non-Gaussian statistics across all source-redshift configurations. MF and MOM give very similar constraining power and generally provide the largest improvements. WST is also competitive, particularly for $w_0$, while peak and minima counts tend to provide smaller gains. The similar performance of MF and MOM is qualitatively consistent with Ref.~\cite{Petri:2013ffb}, which found that MF can capture all the information contained in the MOM if the bin-widths are sufficiently narrow. Combining all five statistics gives tighter constraints than any individual statistic for every source-redshift configuration, indicating that the different summaries contain complementary information.

\begin{table*}
  \centering
  \setlength{\tabcolsep}{3.5pt}
  \renewcommand{\arraystretch}{1.1}
  \small

  \begin{tabular}{lcccccccccccc}
    \hline
    & \multicolumn{4}{c}{$z_s=[0.5,1,1.5,2,2.5,1100]$}
    & \multicolumn{4}{c}{$z_s=[1,2,2.5]$}
    & \multicolumn{4}{c}{$z_s=[1,2.5]$} \\
    \cline{2-5}\cline{6-9}\cline{10-13}
    Statistics
    & $\Omega_m$ & $\sigma_8$ & $w_0$ & $w_a$
    & $\Omega_m$ & $\sigma_8$ & $w_0$ & $w_a$
    & $\Omega_m$ & $\sigma_8$ & $w_0$ & $w_a$ \\
    \hline
    MF  & 34.4\% & 39.4\% & 15.2\% & 20.7\%
        & 28.2\% & 44.1\% & 11.1\% & 34.8\%
        & 23.1\% & 41.1\% & 12.7\% & 34.5\% \\
    PC  & 24.5\% & 28.2\% &  6.9\% & 11.8\%
        & 20.4\% & 35.0\% &  2.6\% & 23.0\%
        & 14.1\% & 31.5\% &  2.7\% & 19.5\% \\
    MC  & 13.8\% & 17.6\% &  4.6\% &  9.8\%
        & 13.4\% & 25.9\% &  2.7\% & 21.3\%
        &  8.5\% & 21.9\% &  3.7\% & 19.2\% \\
    MOM & 35.3\% & 41.2\% & 15.4\% & 21.5\%
        & 26.1\% & 44.2\% &  8.0\% & 35.1\%
        & 20.8\% & 41.7\% & 10.0\% & 35.5\% \\
    WST & 28.9\% & 31.7\% & 16.0\% & 18.6\%
        & 26.6\% & 37.6\% & 12.7\% & 27.7\%
        & 19.2\% & 32.5\% & 11.8\% & 24.0\% \\
    ALL & 41.9\% & 45.8\% & 24.3\% & 28.0\%
        & 35.6\% & 49.6\% & 19.7\% & 40.0\%
        & 28.9\% & 45.4\% & 19.7\% & 39.8\% \\
    \hline
  \end{tabular}

  \vspace{0.8em}

  \begin{tabular}{lcccccccc}
    \hline
    & \multicolumn{4}{c}{$z_s=[1,2.5,1100]$}
    & \multicolumn{4}{c}{$z_s=[2.5,1100]$} \\
    \cline{2-5}\cline{6-9}
    Statistics
    & $\Omega_m$ & $\sigma_8$ & $w_0$ & $w_a$
    & $\Omega_m$ & $\sigma_8$ & $w_0$ & $w_a$ \\
    \hline
    MF  & 26.2\% & 29.6\% & 13.9\% & 19.2\%
        & 27.3\% & 28.4\% & 2.5\% & 2.2\% \\
    PC  & 13.9\% & 16.7\% &  4.2\% &  8.2\%
        & 13.4\% & 14.4\% & 0.9\% & 1.0\% \\
    MC  &  6.2\% &  8.0\% &  3.1\% &  7.3\%
        &  4.2\% &  4.9\% & 1.0\% & 0.6\% \\
    MOM & 26.6\% & 31.1\% & 13.4\% & 19.5\%
        & 28.8\% & 31.2\% & 1.6\% & 1.6\% \\
    WST & 17.7\% & 19.3\% & 11.6\% & 14.8\%
        & 14.2\% & 14.2\% & 1.1\% & 0.6\% \\
    ALL & 32.0\% & 34.6\% & 21.0\% & 26.3\%
        & 32.4\% & 32.9\% & 3.9\% & 3.8\% \\
    \hline
  \end{tabular}

  \caption{Fractional reduction in the marginalized $1\sigma$ uncertainties relative to the power-spectrum-only constraints, $1-\sigma_{C_\ell+X}/\sigma_{C_\ell}$, for each non-Gaussian statistic and source-redshift combination. Here $z_s=1100$ denotes CMB lensing. ALL denotes the joint combination of MF, PC, MC, MOM, and WST.}
  \label{tab:pk_improvement_allzs}
\end{table*}

The comparison between $z_s=[1,2.5]$ and $z_s=[1,2.5,1100]$ illustrates how adding CMB lensing changes the relative contribution of the non-Gaussian statistics. The final constraints become tighter when CMB lensing is included, but the power-spectrum-only baseline also changes. As a result, the fractional gain from non-Gaussian statistics does not change in the same way for every parameter. For example, the improvement in $\sigma(\sigma_8)$ decreases from $45.4\%$ to $34.6\%$, and that in $\sigma(w_a)$ decreases from $39.8\%$ to $26.3\%$, whereas the improvements in $\sigma(\Omega_m)$ and $\sigma(w_0)$ increase slightly. This behavior reflects the fact that CMB lensing already provides additional constraints on the matter distribution and changes the parameter degeneracies present in the galaxy-lensing analysis.

\begin{figure*}
    \centering
    \includegraphics[width=1.0\linewidth]{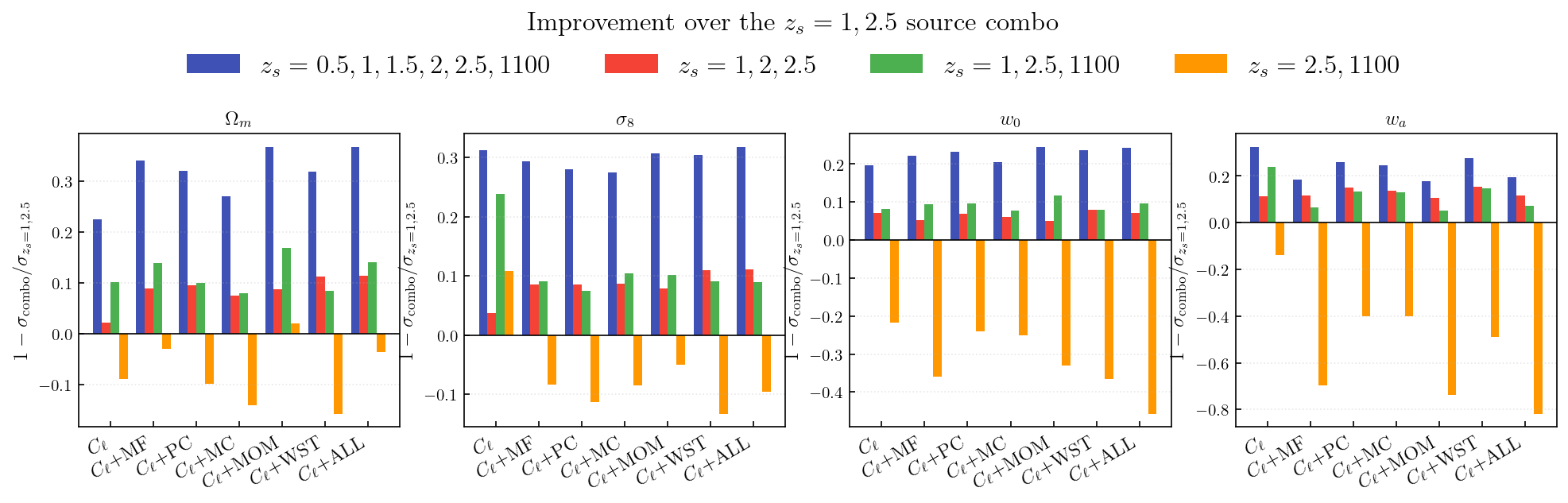}
    \caption{Fractional reduction in marginalized $1\sigma$ uncertainties obtained by changing the source-redshift combination relative to the baseline $z_s=[1,2.5]$: $1-\sigma_{\rm combo}/\sigma_{z_s=[1,2.5]}$, evaluated separately for each data vector. Panels show $\Omega_m$, $\sigma_8$, $w_0$, and $w_a$ from left to right. Within each panel, the groups correspond to $C_\ell$ alone and $C_\ell$ combined with Minkowski functionals (MF), peak counts (PC), minima counts (MC), moments (MOM), the wavelet scattering transform (WST), or all five jointly ($C_\ell+\mathrm{ALL}$). Bar colors indicate the alternative source-redshift combinations, with $z_s=1100$ denoting CMB lensing.}
    \label{fig:forecast_improve_10_25_barplot}
\end{figure*}

Figure~\ref{fig:forecast_improve_10_25_barplot} shows more directly how the constraints change when the source-redshift coverage is increased relative to $z_s=[1,2.5]$. Adding either the intermediate galaxy source plane at $z_s=2$ or CMB lensing improves the constraints even when only the power spectra are used. The tightest constraints are obtained when all available galaxy source planes and CMB lensing are combined, as expected. This demonstrates that extending the redshift coverage provides information that is complementary to the non-Gaussian information contained within each lensing convergence field.

Overall, tomography and non-Gaussian statistics play distinct but complementary roles. Tomography is particularly important for breaking degeneracies involving the dark-energy parameters, while non-Gaussian statistics recover additional information from nonlinear structure that is not captured by the power spectrum. Combining both therefore provides the strongest constraints in the four-dimensional parameter space considered here.

If we assume that the current DESI+DES best-fit values of $w_0$ and $w_a$ remain unchanged, we can also estimate how the increased precision of our forecast would affect the detection significance of the departure from $\Lambda$CDM. The DESI+DES combination currently finds a $3.0\sigma$ deviation from $(w_0,w_a)=(-1,0)$~\cite{DES:2026jmi}. Approximating both the DESI+DES and forecast posteriors as Gaussian, and assuming that the $w_0$--$w_a$ correlation remains unchanged, the reduced marginalized uncertainties obtained when combining all source redshifts with Rubin-like and SO-like noise would increase the significance to approximately $5.2\sigma$ when all non-Gaussian statistics are included, compared with $3.8\sigma$ using the tomographic power spectra alone. This estimate should be interpreted as illustrative and optimistic, since the exact significance depends on the full covariance and on whether the central values and degeneracy direction remain stable as the data improve. Moreover, this is not a direct comparison with the DESI+DES constraints, as our analysis varies only four cosmological parameters and does not include observational systematics.

\subsection{Contributions of Individual Minkowski Functionals and Moments}

\begin{figure*}
    \centering
    \includegraphics[width=1.0\linewidth]{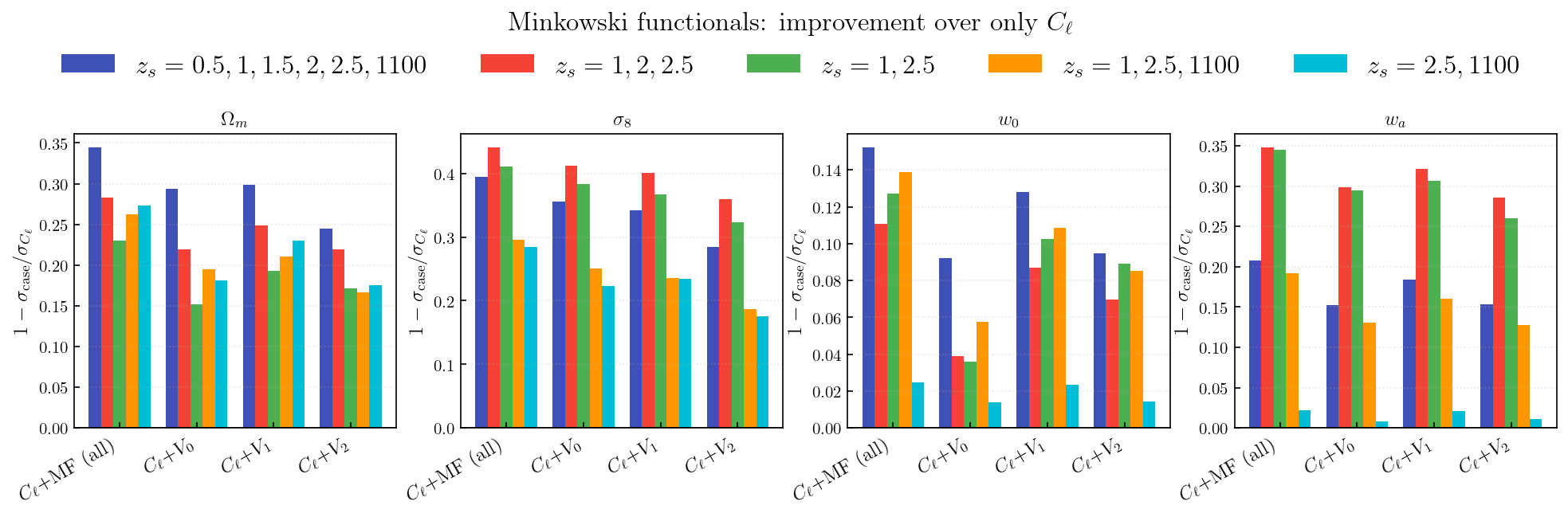}
    \includegraphics[width=1.0\linewidth]{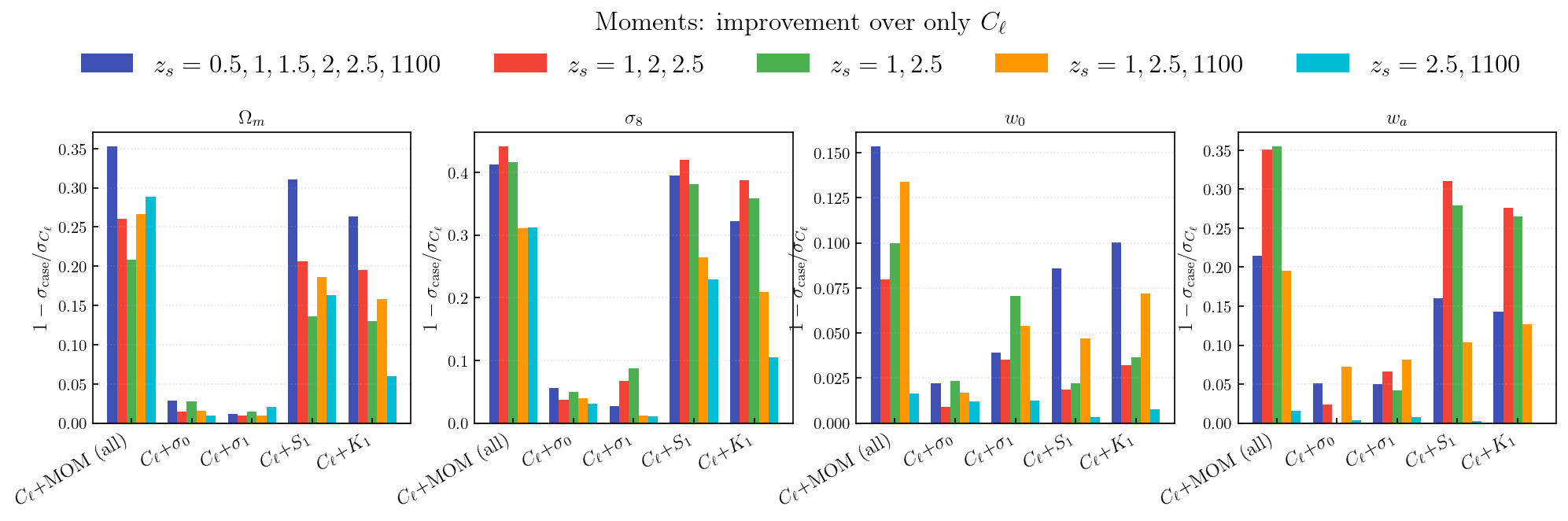}
    \caption{Fractional reduction in marginalized $1\sigma$ uncertainties obtained by adding individual Minkowski functionals (upper panels) or moments (lower panels) to the lensing power spectra, $1-\sigma_{\rm case}/\sigma_{C_\ell}$. Each row shows $\Omega_m$, $\sigma_8$, $w_0$, and $w_a$ from left to right. Within each panel, the groups correspond to the added functional or moment, and bar colors indicate the source-redshift combinations, with $z_s=1100$ denoting CMB lensing.}
    \label{fig:forecast_improveCell_barplot_mfmom}
\end{figure*}

The combined MF and MOM constraints do not reveal which individual components are responsible for the additional information beyond the power spectrum. We therefore repeat the analysis by adding each Minkowski functional or moment separately to the power-spectrum data vector. Figure~\ref{fig:forecast_improveCell_barplot_mfmom} shows the resulting fractional reductions in the marginalized parameter uncertainties for the different source-redshift combinations.

The three Minkowski functionals contribute differently depending on the parameter being constrained. For $\sigma_8$, $V_0$ provides the largest improvement, followed by $V_1$ and $V_2$. For $\Omega_m$, $w_0$, and $w_a$, the ordering is reversed, with $V_1$ providing the largest gain, followed by $V_2$ and then $V_0$. This difference reflects the distinct information encoded by the three functionals. While $V_0$ is closely related to the one-point distribution of the convergence field, $V_1$ and $V_2$ are additionally sensitive to the geometry and topology of excursion regions. The parameter dependence of the ranking therefore suggests that amplitude information and morphological information contribute differently to the cosmological constraints.

A similar decomposition can be performed for the moments. The lower-order quantities $\sigma_0$ and $\sigma_1$ generally provide relatively small improvements, although $\sigma_1$ contributes more noticeably to the $w_0$ constraint. Most of the constraining power beyond the power spectrum is instead carried by the higher-order derivative moments $S_1$ and $K_1$. The performance of the full MOM data vector is therefore driven primarily by these higher-order quantities rather than by the rms convergence and gradient amplitudes alone. This also helps explain why the moment data vector can recover information that is absent from the power spectrum.

\subsection{Comparison with the Prior and Current Constraints}

We next compare the constraining power of our full analysis with both the priors adopted in the inference and current observational constraints. Table~\ref{tab:prior_improvement_all} shows the fractional reduction in the marginalized $1\sigma$ uncertainties relative to the Gaussian priors for $C_\ell+\mathrm{ALL}$, where all five non-Gaussian statistics are combined with the power spectrum. Using all available galaxy source redshifts together with CMB lensing, the posterior uncertainties on $\Omega_m$, $\sigma_8$, $w_0$, and $w_a$ are reduced by $70.5\%$, $85.0\%$, $67.4\%$, and $66.7\%$ compared to the priors, respectively. Thus, even with the relatively informative priors adopted here, the lensing data provide substantial additional constraints on all four parameters. In particular, the uncertainties on both dark-energy parameters are reduced by approximately two thirds relative to their prior widths.

We also compare our forecast with constraints from the full DES analysis~\cite{DES:2026jmi}. Table~\ref{tab:ref_improvement_all} considers two reference cases: the DES combination excluding the $3\times2$pt measurements and the full DES data combination. Relative to the full DES constraint, our configuration with all galaxy source planes and CMB lensing gives $40.9\%$, $62.5\%$, $34.7\%$, and $48.7\%$ smaller marginalized uncertainties on $\Omega_m$, $\sigma_8$, $w_0$, and $w_a$, respectively. Relative to the DES combination without $3\times2$pt information, the corresponding reductions are $46.3\%$, $68.4\%$, $40.7\%$, and $52.4\%$.

These numbers demonstrate the statistical potential of combining tomographic galaxy and CMB lensing with information beyond the power spectrum, but they should not be interpreted as a direct like-for-like comparison with DES. The two analyses differ in survey specifications, observables, scale ranges, and treatment of systematic effects. Our forecast, for example, does not include the galaxy overdensity field, and the finite size of the simulated maps removes the largest-scale modes, both of which could provide additional cosmological information (indeed, current CMB lensing maps are only signal-dominated on angular scales larger than those included in our maps). On the other hand, we neglect several observational and astrophysical systematics that are included in real-data analyses. The comparison with DES is therefore intended primarily to place the statistical precision of our forecasts in the context of current constraints.

The importance of redshift coverage is also apparent in Table~\ref{tab:ref_improvement_all}. Configurations containing multiple galaxy source planes generally yield tighter constraints than the DES reference values for all four parameters. In contrast, the $z_s=[2.5,1100]$ configuration, which contains only one galaxy source plane together with CMB lensing, gives weaker constraints on $w_0$ and $w_a$ than DES, as indicated by the negative fractional improvements. This again highlights the importance of low- and intermediate-redshift galaxy-lensing information for constraining the evolution of dark energy.

\begin{table}[t]
  \centering
  \begin{tabular}{lcccc}
    \hline
    Source redshifts & $\Omega_m$ & $\sigma_8$ & $w_0$ & $w_a$ \\
    $\sigma_{\mathrm{prior}}$ & 0.01 & 0.02 & 0.1 & 0.3 \\
    \hline
    $0.5,1,1.5,2,2.5,1100$ & 70.5\% & 85.0\% & 67.4\% & 66.7\% \\
    $1,2,2.5$               & 58.7\% & 80.4\% & 59.9\% & 63.4\% \\
    $1,2.5$                 & 53.4\% & 78.0\% & 56.8\% & 58.7\% \\
    $1,2.5,1100$            & 60.0\% & 80.0\% & 61.0\% & 61.6\% \\
    $2.5,1100$              & 51.7\% & 75.9\% & 37.1\% & 24.9\% \\
    \hline
  \end{tabular}
  \caption{Fractional reduction in the marginalized $1\sigma$ parameter uncertainties relative to the adopted Gaussian priors (see Eq.~\eqref{eq.priors}), $1-\sigma_{\mathrm{post}}/\sigma_{\mathrm{prior}}$, for $C_\ell+\mathrm{ALL}$. Here ALL denotes the joint combination of MF, PC, MC, MOM, and WST. Results are shown for each source-redshift combination, with $z_s=1100$ denoting CMB lensing.}
  \label{tab:prior_improvement_all}
\end{table}

\begin{table}[t]
  \centering
  \begin{tabular}{lcccc}
    \hline
    Source redshifts & $\Omega_m$ & $\sigma_8$ & $w_0$ & $w_a$ \\
    \hline
    $\sigma_{\mathrm{ref}}$ -- DESI+DES w/o $3\times2$pt
      & 0.0055 & 0.0095 & 0.055 & 0.21 \\
    \hline
    $0.5,1,1.5,2,2.5,1100$ & 46.3\% & 68.4\% & 40.7\% & 52.4\% \\
    $1,2,2.5$               & 25.0\% & 58.8\% & 27.1\% & 47.8\% \\
    $1,2.5$                 & 15.3\% & 53.7\% & 21.5\% & 40.9\% \\
    $1,2.5,1100$            & 27.2\% & 57.8\% & 29.1\% & 45.2\% \\
    $2.5,1100$              & 12.2\% & 49.3\% & -14.3\% & -7.4\% \\
    \hline
    $\sigma_{\mathrm{ref}}$ -- Full DESI+DES
      & 0.005 & 0.008 & 0.05 & 0.195 \\
    \hline
    $0.5,1,1.5,2,2.5,1100$ & 40.9\% & 62.5\% & 34.7\% & 48.7\% \\
    $1,2,2.5$               & 17.5\% & 51.1\% & 19.8\% & 43.8\% \\
    $1,2.5$                 & 6.8\%  & 45.0\% & 13.7\% & 36.4\% \\
    $1,2.5,1100$            & 19.9\% & 49.9\% & 22.0\% & 40.9\% \\
    $2.5,1100$              & 3.4\%  & 39.8\% & -25.8\% & -15.6\% \\
    \hline
  \end{tabular}
  \caption{Fractional reduction in the marginalized $1\sigma$ parameter uncertainties from $C_\ell+\mathrm{ALL}$ relative to two reference parameter constraint sets from the full DESI+DES analysis~\cite{DES:2026jmi}, $1-\sigma_{\mathrm{post}}/\sigma_{\mathrm{ref}}$. The first reference excludes the $3\times2$pt information, while the second uses the full DESI+DES data combination. Positive values indicate tighter constraints in our forecast, while negative values indicate weaker constraints. Here $z_s=1100$ denotes CMB lensing.}
  \label{tab:ref_improvement_all}
\end{table}

\section{Conclusions}\label{sec:conclusions}

Using {\tt DDELenS}~\cite{Chen:2026zqq}, a suite of fully ray-traced lensing convergence maps generated from $N$-body simulations, we have investigated the constraining power of non-Gaussian lensing statistics for dynamical dark energy. We considered Minkowski functionals, moments, peak counts, minima counts, and the wavelet scattering transform for Rubin-like galaxy lensing and Simons Observatory-like CMB lensing configurations. By combining these statistics with tomographic power spectra across several source-redshift configurations, we quantified how much information beyond two-point statistics remains available once redshift information is included.

For configurations with multiple galaxy-lensing source planes, non-Gaussian statistics provide a substantial improvement over the power-spectrum baseline. Combining all five statistics reduces the marginalized uncertainties on $w_0$ and $w_a$ by roughly $20\%$ and $40\%$, respectively, for representative tomographic configurations, while also significantly improving the constraints on $\Omega_m$ and $\sigma_8$. The gains persist when all available source redshifts are included, demonstrating that non-Gaussian information remains complementary to lensing tomography rather than becoming redundant once the redshift baseline is extended. This is consistent with previous studies showing that combining tomographic information with non-Gaussian lensing observables can substantially improve cosmological and dark-energy constraints.

Among the individual summary statistics, Minkowski functionals and moments provide comparable constraining power and generally yield the largest gains, with the wavelet scattering transform also performing competitively. Decomposing the moments shows that much of their additional information is carried by the higher-order derivative moments $S_1$ and $K_1$. The fact that the joint data vector consistently outperforms any individual statistic further indicates that the different summaries probe complementary aspects of the nonlinear lensing field. 

Our results also highlight the importance of combining non-Gaussian information with broad redshift coverage. CMB lensing adds a high-redshift projection of the matter distribution and changes the parameter degeneracies present in galaxy lensing alone, while multiple galaxy source planes are essential for resolving the redshift evolution associated with $w_0$ and $w_a$. The complementarity between galaxy surveys and CMB lensing has also been demonstrated in joint forecasts for Rubin Observatory and the Simons Observatory~\cite{Fang:2021ici}. The weak improvement obtained from non-Gaussian statistics when only a single galaxy source plane is combined with CMB lensing therefore reflects the limited dark-energy leverage of that configuration, and a lack of non-Gaussian information in the lensing field itself for high-redshift source planes where the density field is more linear. The strongest constraints arise when tomography and non-Gaussian observables are used together.

The forecasts presented here are intentionally idealized. We neglect baryonic effects, intrinsic alignments, and realistic source-redshift distributions, and vary only a limited set of cosmological parameters. These ingredients will need to be incorporated before applying the framework to survey data. Their importance for non-Gaussian lensing analyses has already been demonstrated in both simulation studies and analyses of current survey data~\cite{DES:2024jgw,DES:2026vyi}. At the same time, they provide clear directions for extending the analysis. Recent work suggests that higher-order statistics can retain useful cosmological information even in the presence of baryonic effects~\cite{Tersenov:2026vfq}, and extending the parameter space will allow us to test how robust the non-Gaussian gains remain in the presence of additional cosmological degeneracies~\cite{Broxterman:2026ovq} (see Ref.~\cite{Vadala:2026gfm} for how higher-order statistics can help distinguish between modified gravity and massive neutrinos).

Taken together, our results show that non-Gaussian lensing statistics have the potential to become an important component of dynamical-dark-energy analyses with upcoming weak-lensing and CMB surveys. As the statistical precision of Rubin, Euclid, Roman, and the Simons Observatory continues to improve, extracting information beyond the power spectrum will become increasingly valuable. The combination of tomographic galaxy lensing, CMB lensing, and complementary non-Gaussian observables therefore provides a promising route toward more stringent and independent tests of time-dependent dark energy. It is also promising to significantly increase our sensitivity to detect departures from a constant vacuum energy density.

\section{Acknowledgments}
SFC, JCH, and ZH acknowledge support from NASA grant 80NSSC24K1093 [ATP]. This work was performed in part at Aspen Center for Physics, which is supported by National Science Foundation grant PHY-2210452. The authors also acknowledge the Texas Advanced Computing Center (TACC) at The University of Texas at Austin for providing computational resources that have contributed to the research results reported within this paper. We acknowledge computing resources from Columbia University’s Shared Research Computing Facility project, which is supported by NIH Research Facility Improvement Grant 1G20RR030893-01, and associated funds from the New York State Empire State Development, Division of Science Technology and Innovation (NYSTAR) Contract C090171, both awarded April 15, 2010.  This is not an official Simons Observatory collaboration paper. The simulation data will be made publicly available upon acceptance of this manuscript.


\appendix
\section{Covariance Matrix of the Gaussian Likelihood}\label{appendix:correlation}
\subsection{Correlation Matrix for Selected Source Redshifts}

\begin{figure*}
  \centering
  \includegraphics[width=0.8\linewidth]{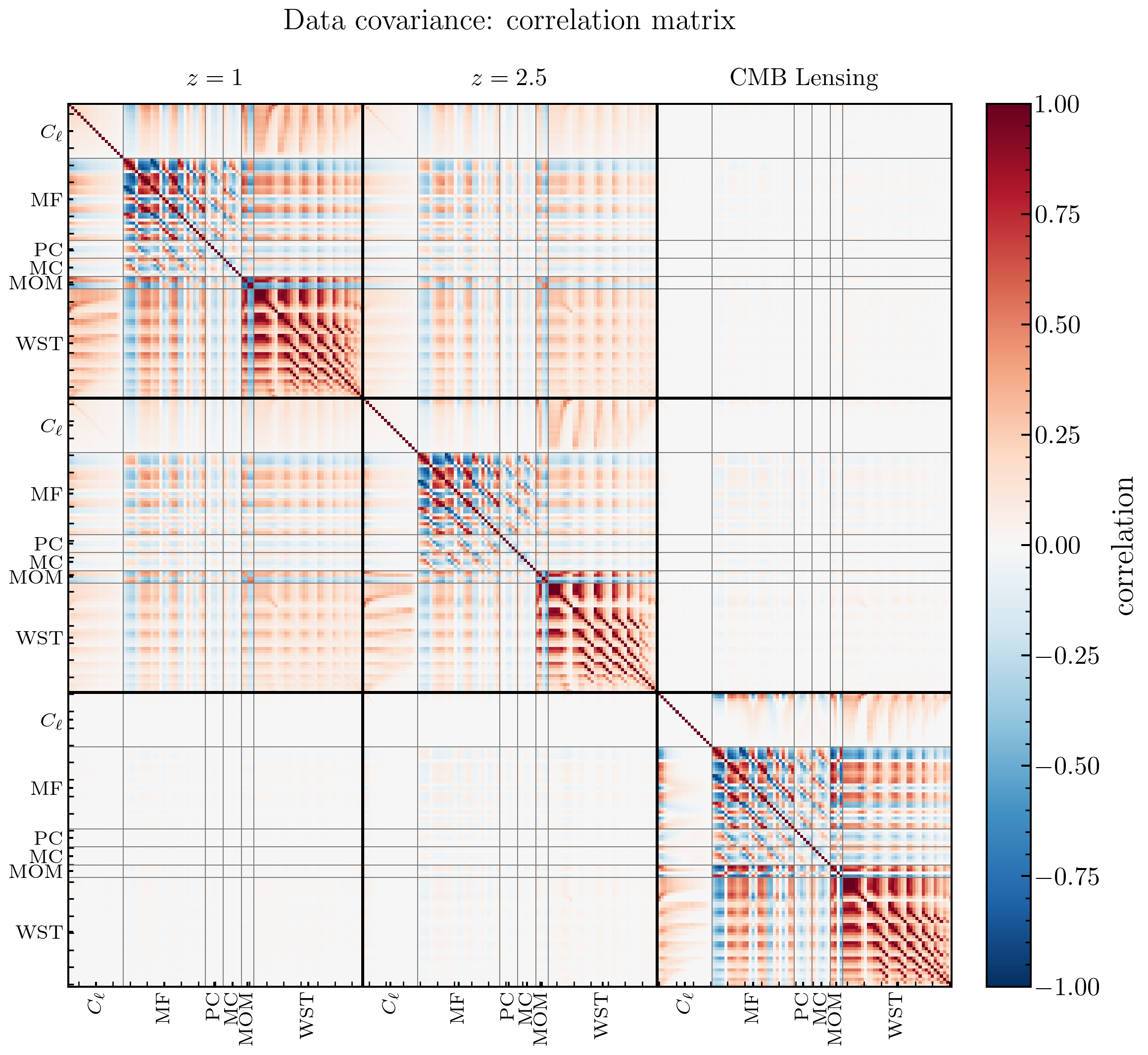}
  \caption{Correlation matrix of the joint data vector for $z_s=1$, $2.5$, and $1100$. Each block corresponds to one source-plane--statistic pair, ordered by source redshift. Thin gray lines separate the different statistics within each source plane, while thick black lines separate the three source planes, labeled along the top. Within each source plane, the statistics are ordered as the power spectrum $C_\ell$, Minkowski functionals ($V_0$, $V_1$, $V_2$), peak counts, minima counts, moments, and wavelet scattering transform coefficients, restricted to the features included in the likelihood. Here $z_s=1100$ denotes CMB lensing.}
  \label{fig:covariance}
\end{figure*}

Figure~\ref{fig:covariance} shows the correlation matrix of the joint data vector for three source redshifts, $z_s=1$, $2.5$, and $1100$. The different statistics measured from the same source plane exhibit substantial correlations, as expected since they are extracted from the same underlying convergence field. We also find clear correlations between the statistics measured at $z_s=1$ and $z_s=2.5$, reflecting the large overlap in the structures contributing to the two galaxy-lensing maps. In comparison, the cross-correlations between the galaxy-lensing statistics and those from CMB lensing are generally weaker. This is consistent with the broader and higher-redshift sensitivity of the CMB-lensing kernel, which contains contributions from structures not probed by the galaxy-lensing source planes. The nonzero cross-correlations nevertheless show that galaxy and CMB lensing retain shared information about the intervening large-scale structure.

\subsection{Convergence of the Covariance Matrix}
\begin{figure*}
    \centering
    \includegraphics[width=1.0\linewidth]{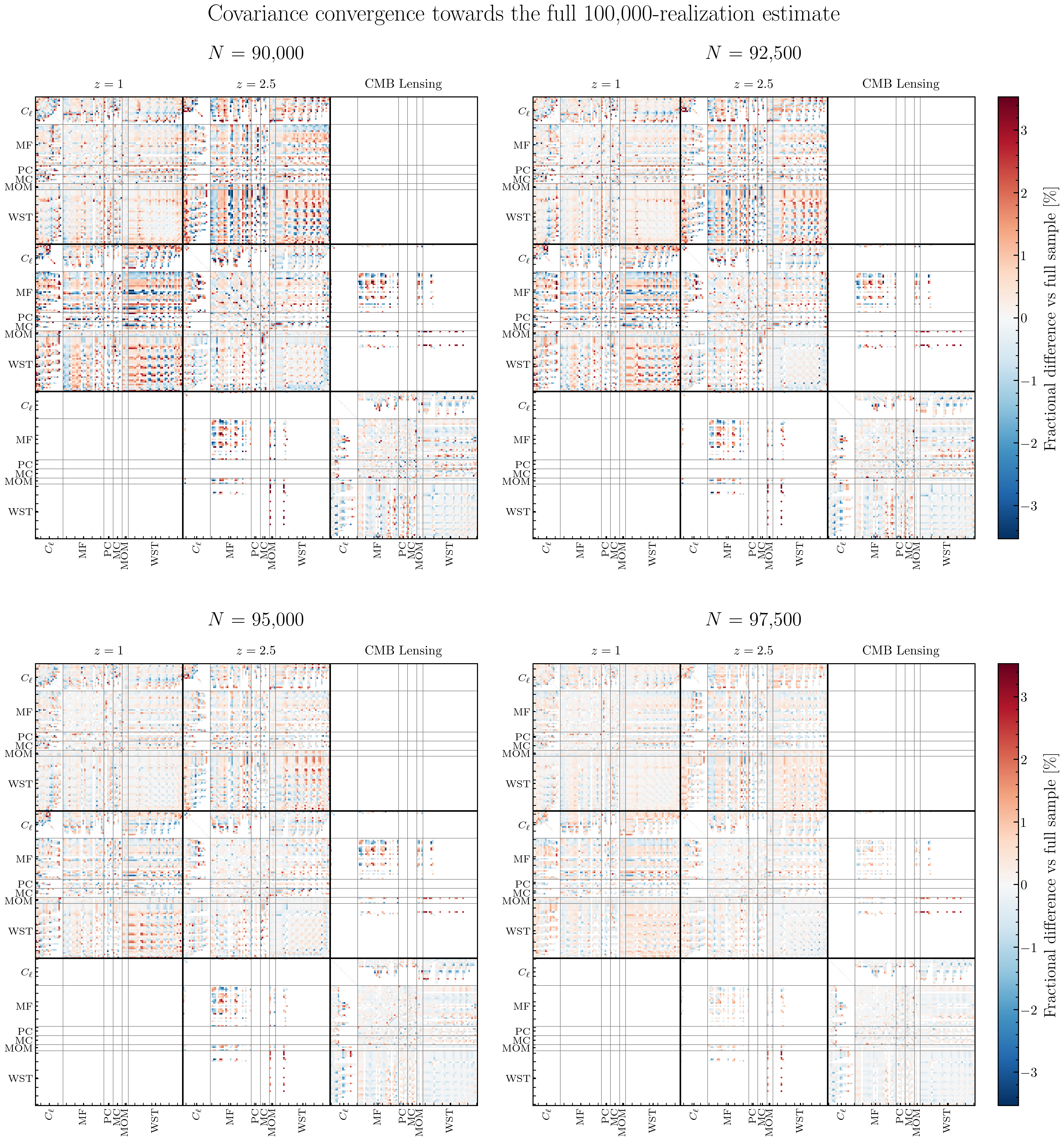}
    \caption{Convergence of the covariance matrix estimated from the mock realizations for statistics from selected source redshifts. We show the fractional difference between the covariance estimated from the full set of 100,000 realizations and estimates obtained from nested subsets of 90,000, 92,500, 95,000, and 97,500 realizations. Elements with correlation coefficients $|\rho_{ij}|<0.05$ are set to zero, since fractional differences become poorly defined for covariance elements that are close to zero.}
    \label{fig:covariance_convergence}
\end{figure*}

We verify that the $N_{\rm ref}=100{,}000$ mock realizations used to estimate the data covariance are sufficient for the covariance to converge. We recompute the covariance using subsets of the first $N\in\{90{,}000,92{,}500,95{,}000,97{,}500\}$ realizations and compare each estimate with the full covariance through the fractional difference $\Delta_{ij}=[C_{ij}(N)-C_{ij}(N_{\rm ref})]/|C_{ij}(N_{\rm ref})|$, as shown in Figure~\ref{fig:covariance_convergence}. The expected root-mean-square fractional difference of a variance is $\sqrt{2(1/N-1/N_{\rm ref})}$ for a Wishart-distributed covariance estimator. The diagonal elements differ from the full estimate by a median of $0.11\%$ at $N=90{,}000$ and $0.06\%$ at $N=97{,}500$, consistent with the expected values of $0.15\%$ and $0.07\%$, respectively. This indicates that the diagonal elements have converged at the level expected from covariance-estimation noise.

The off-diagonal elements show larger fractional differences, as expected because the relative error of $C_{ij}$ scales as $\sqrt{(1+\rho_{ij}^{-2})/N}$ and therefore becomes larger for weakly correlated pairs. In this case, the absolute change in the correlation coefficient provides a more useful measure of convergence. We find $\max_{ij}|\Delta\rho_{ij}|\leq0.005$ even at $N=90{,}000$. We therefore conclude that the covariance has converged and that the residual covariance-estimation noise is negligible for the constraints.

\clearpage
\newpage

\bibliography{ref.bib}
\end{document}